\documentclass[prb,twocolumn,showpacs,preprintnumbers]{revtex4}

\usepackage{graphicx}
\usepackage{dcolumn}
\usepackage{bm}

\begin{document}

\preprint{APS/123-QED}

\title{Ga-induced atom wire formation and passivation of stepped
Si(112)}

\author{P.C. Snijders}
 \email{p.c.snijders@tnw.tudelft.nl}
\affiliation{Kavli Institute of NanoScience Delft, Delft
University of Technology, 2628 CJ Delft, The Netherlands}

\author{C. Gonz\'{a}lez} \affiliation{Facultad de
Ciencias, Departamento de F\'{i}sica Te\'{o}rica de la Materia
Condensada, Universidad Aut\'{o}noma de Madrid, Madrid 28049,
Spain}

\author{S. Rogge}
\affiliation{Kavli Institute of NanoScience Delft, Delft
University of Technology, 2628 CJ Delft, The Netherlands}

\author{R. P\'{e}rez}
\affiliation{Facultad de Ciencias, Departamento de F\'{i}sica
Te\'{o}rica de la Materia Condensada, Universidad Aut\'{o}noma de
Madrid, Madrid 28049, Spain}

\author{J. Ortega}
\affiliation{Facultad de Ciencias, Departamento de F\'{i}sica
Te\'{o}rica de la Materia Condensada, Universidad Aut\'{o}noma de
Madrid, Madrid 28049, Spain}

\author{F. Flores}
\affiliation{Facultad de Ciencias, Departamento de F\'{i}sica
Te\'{o}rica de la Materia Condensada, Universidad Aut\'{o}noma de
Madrid, Madrid 28049, Spain}

\author{H.H. Weitering}
\affiliation{Department of Physics and Astronomy, The University
of Tennessee, Knoxville, TN 37996, USA, and\\Condensed Matter
Sciences Division, Oak Ridge National Laboratory, Oak Ridge, TN
37831, USA}

\date{\today}

\begin{abstract}
We present an in-depth analysis of the atomic and electronic
structure of the quasi one-dimensional (1D) surface reconstruction
of Ga on Si(112) based on Scanning Tunneling Microscopy and
Spectroscopy (STM and STS), Rutherford Backscattering Spectrometry
(RBS) and Density Functional Theory (DFT) calculations. A new
structural model of the Si(112)$6\times1$-Ga surface is inferred.
It consists of Ga \emph{zig-zag} chains that are intersected by
quasi-periodic vacancy lines or misfit dislocations. The
experimentally observed meandering of the vacancy lines is caused
by the co-existence of competing $6\times1$ and $5\times1$ unit
cells and by the orientational disorder of symmetry breaking Si-Ga
dimers inside the vacancy lines. The Ga atoms are fully
coordinated, and the surface is chemically passivated. STS data
reveal a semiconducting surface and show excellent agreement with
calculated Local Density of States (LDOS) and STS curves. The
energy gain obtained by fully passivating the surface calls the
idea of step-edge decoration as a viable growth method toward 1D
metallic structures into question.
\end{abstract}

\maketitle

\section{\label{sec:intro}Introduction\protect}
Nature only provides few one-dimensional (1D) electronic systems,
such as carbon nanotubes,\cite{Dekker:00} organic charge transfer
salts, and inorganic blue bronzes (see for example the discussion
in Ref.~\onlinecite{Grioni:99}). Electrons confined to one
dimension are fundamentally different from the quasi-particles of
Fermi liquid theory.\cite{Voit:94} In 1D, even in the case of
arbitrary low interaction strength, the single-particle
description of the system breaks down and must be replaced by a
description based on collective excitations.\cite{Voit:94}
Experimental realization and verification of this Luttinger liquid
phenomenon continues to capture the imagination of physicists,
especially since the fabrication of structurally uniform 1D
nanostructures now appears to be within the realm of
possibilities.

A very intuitive approach to produce 1D systems is to utilize high
index silicon surfaces.\cite{Himpsel:01} Based on the concept of
metal-adatom step-edge decoration, deposition of a submonolayer
amount of metal atoms onto a stepped Si surface is expected to
result in a \emph{single domain} of quasi 1D, metallic
\emph{atomic} wires, \emph{i.e.} an atom wire
array.\cite{Himpsel:01} In contrast to for example carbon
nanotubes, such a single domain surface quantum wire array would
be easily accessible to both nanoscopic and macroscopic techniques
such as Scanning Tunneling Microscopy and Spectroscopy (STM and
STS), photoemission spectroscopy, and (surface) transport
measurements. In addition, the coupling strength between the atom
wires can be tuned by changing the miscut angle of the vicinal Si
surface, \emph{i.e.} adjusting the separation between the
wires.\cite{Himpsel:04}

Indeed such single domain 1D metallic systems have been produced
on high index Si
surfaces.\cite{Himpsel:04,Baer:02,Baer:99,Yeom:99,Ahn:04} However,
the metal adatoms in these studies are generally \emph{not}
adsorbed at the step edges. Instead, rather complicated
reconstructions are formed with chains of metal atoms that are
incorporated into the (111)-like terraces in the unit
cell.\cite{Bunk:99,Himpsel:02,Himpsel:04} This questions the idea
of forming atom wires via step-edge decoration on Si. Furthermore,
it was noticed\cite{Gonzalez:04} that these 1D atomic-scale
systems all exhibit intrinsic spatial disorder in the atomic
structure, which will have important consequences for electronic
transport in these systems. It should be noted that in the case
that step-edge decoration does not occur in these studies, the
miscut or vicinal orientation of the Si surfaces mainly serves to
create a \emph{single domain} surface reconstruction; similar or
"parent" reconstructions exist on the corresponding planar
surfaces. These single domain quantum wire arrays have been
studied successfully with Angle Resolved Photoelectron
Spectroscopy
(ARPES)\cite{Crain:03,Himpsel:04,Ahn:03,Ahn:04,Matsuda:03} and
transport measurements.\cite{Hasegawa:03}

\begin{figure}[ht]
  \centering
  \includegraphics[width=\columnwidth]{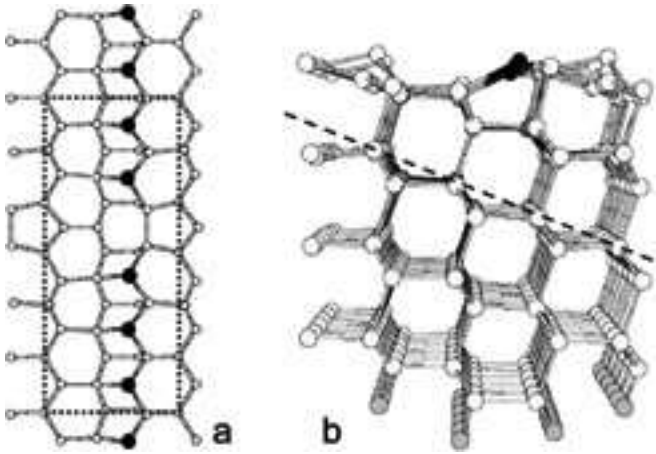}
  \caption{(\emph{a}) Topview and (\emph{b}) side view of the
  JKP-model of the Si(112)$6\times1$-Ga surface. In
  (\emph{a}) the $6\times1$ unit cell is indicated dotted. In (\emph{b}) a
  (111) plane is indicated. Si atoms: light, Ga atoms: dark.}
  \label{fig:structold}
\end{figure}

In this paper, we investigate the formation of Ga chains on the
vicinal Si(112) surface. A structural model for this interface was
devised by Jung, Kaplan and Prokes (the
''JKP-model'').\cite{Jung:93,Jung:94,Yater:95,Baski:99,Erwin:99,Yoo:02}
The unit cell of the bulk terminated vicinal Si(112) surface
contains a double-width (111)-like terrace with single (111)-like
steps. Based on Low Energy Electron Diffraction (LEED) and Auger
Electron Spectroscopy (AES)
experiments,\cite{Jung:93,Jung:94,Yater:95} it was proposed that
Ga atoms adsorb at the step edges of the bulk terminated unit
cell, thus forming atom rows along the [1$\overline{1}$0]
direction. Missing Ga atoms or vacancies in these Ga rows align
into quasi 1D vacancy lines that run orthogonal to the Ga rows,
resulting in the observed $6\times1$ periodicity. In this model
the Ga coverage is $\frac{1}{6}$ of a Si(111) bilayer, or five
atoms per $6\times1$ unit cell.\cite{Jung:94} Later STM
experiments by Baski \emph{et al.}\cite{Baski:96,Baski:99} seemed
to confirm this model. These authors observed a well-ordered array
of single-atom rows with a regular row spacing 9.4~\AA, equal to
the step-edge spacing of bulk terminated Si(112); see for example
Fig. 1 of Ref.~\onlinecite{Baski:99}. A side view and a top view
representation of this JKP-model is shown in
Fig.~\ref{fig:structold}.

As a consequence of the three-fold coordination of the adsorption
sites, the trivalent Ga atoms are fully coordinated. There are no
partially filled dangling bonds on the Ga atoms and the covalently
bonded Ga atoms would not contribute any state density near the
Fermi level. However, an interesting feature which has remained
largely unnoticed in literature is the fact that within the JKP
model, there should exist a metallic dangling bond wire that is
located on the row of Si surface atoms located in between the Ga
rows. However, the predicted 1D metallicity turned out to be
unstable with respect to a Jahn-Teller distortion, leaving only
one unpaired electron per $6\times1$ unit cell. Interestingly, the
resulting electronic structure implied the existence of conduction
channels \emph{orthogonal} to the Ga chains.\cite{Yoo:02}

We have performed a detailed Scanning Tunneling Microscopy (STM)
study of the Si(112)$6\times1$-Ga surface. Because of the
unprecedented resolution in the STM images of the
Si(112)$6\times1$-Ga surface, a detailed investigation of the
atomic structure of the Si(112)$6\times1$-Ga surface could be
carried out. Extensive Density Functional Theory (DFT)
calculations have been performed to explore new candidate
structural models. Theoretical STM images were calculated for the
new structures and compared with the experimental STM images. From
a detailed analysis of all experimental and theoretical
information, a new structural model for the Si(112)$6\times1$-Ga
surface emerged, which shows excellent agreement with the
experimental evidence. It contains \emph{two} Ga atom rows
amounting to a total of ten Ga atoms per $6\times1$ unit cell,
consistent with RBS experiments. The two Ga rows form
\emph{zig-zag} chains while quasi-periodic vacancy lines intersect
these Ga chains. The observed meandering of the vacancy lines can
also be fully explained within this model. STS measurements show
that the surface is semiconducting, and are consistent with our
band structure calculations and theoretical STS simulations. This
paper presents a follow-up of an initial report\cite{Gonzalez:04}
with new data and provides a more detailed and in depth analysis,
including a detailed comparison between spatially resolved STS and
theoretical local density of states (LDOS) calculations.

\section{\label{sec:exp}Experimental and theoretical procedures}
Experiments were carried out in an ultra-high vacuum system with a
base pressure $<$~5~$\times~10^{-11}$~mbar. The system was
equipped with a Ga effusion cell, direct current sample heating
facilities, an Omicron variable temperature STM and a LEED system.
An \emph{n}-type Si(112) wafer ($\sim5\times10^{14} cm^{-3}$,
orientation $\pm$~2$^\circ$ of the nominal (112) orientation) was
cut into (10$\times$2)-mm$^{2}$ samples and rinsed in acetone and
isopropanol. After introduction into UHV the samples were degassed
at 775 K overnight and subsequently the sample temperature was
slowly raised to 1025 K and kept there for 4 hrs. Next, the sample
was flashed at 1475 K to remove the native oxide. During resistive
heating, the current was directed parallel to the nano-facets of
the clean (112) surface (\emph{i.e.} in the [1$\overline{1}$0]
direction) in order to avoid current-induced step bunching. The
surface reconstruction was prepared in two different ways. In the
"one-step" procedure, Ga was deposited with the Si substrate held
at 825 $\pm$ 50 K.\cite{Glembocki:97} In the "two-step" procedure,
Ga is deposited onto a Si(112) substrate kept at room temperature.
After Ga deposition the sample was annealed at about 825 $\pm$ 50
K to form the 6$\times$1 reconstruction and to desorb excess Ga
atoms.\cite{Baski:99} Both surface preparation procedures resulted
in identical LEED patterns and STM images. The pressure remained
below 2$\times10^{-10}$ mbar during sample preparation. The sample
temperature during sample preparation was measured using an
optical pyrometer. STM and STS experiments were performed at room
temperature and at low temperature ($\sim 40$ K) using etched
tungsten tips. STM images of the filled and empty electronic
states were obtained with a constant current between 0.05 and
0.2~nA and bias voltages between 1 and 2~V. STS data were acquired
with a setpoint of 0.3 nA at 1 V.

RBS experiments were carried out at the AMOLF institute in
Amsterdam to determine the amount of Ga atoms per surface unit
cell. A normal incident 2.0~MeV He$^+$ ion beam from a Van de
Graaf accelerator was backscattered from the Si crystal and
detected at a backscattering angle of 165$^\circ$. The beam
current was typically about 20~nA.

The atomic and electronic structure of new candidate structural
models for the Si(112)$6\times1$-Ga surface, corresponding to
Ga-coverages ranging from 5 to 11 Ga atoms per 6$\times$1 unit
cell were explored using an efficient local-orbital (LO) DFT
technique (the {\sc Fireball96} code).\cite{Demkov:95} In these
calculations, we have used a minimal atomic-like basis set using
the following cut-off radii ($R_c$) for the definition of the {\sc
Fireball96} orbitals \cite{Sankey:89}: $R_c(Si) = 5.0$, $R_c(Ga) =
5.2$. For the most promising structures, Plane-Waves (PW) DFT
calculations ({\sc Castep} code)\cite{CASTEP} were also performed
to check the validity of the {\sc Fireball96} findings. In these
PW calculations, we have used a kinetic energy cut-off $E_c$ of
200 eV for the definition of the PW basis set, and 4 special
$k$-points for the Brillouin zone sampling (test calculations with
250 eV and 8 special $k$-points were also performed). In both the
LO and PW calculations we have used a slab of 11 Si layers with
hydrogen atoms saturating the bonds of the deeper Si layers (see
Fig.~\ref{fig:structold}).

Using the DFT local-orbital hamiltonian of the surface together
with non-equilibrium Keldysh Green function
techniques,\cite{Mingo:96,Jurczyszyn:01} we calculated theoretical
STM images for the new relaxed atomic structures. The theoretical
images were then compared with the experimental STM images. In our
approach, we divide the total hamiltonian, $\hat{H}$, of our
tip-sample system into three parts, $\hat{H} = \hat{H}_t +
\hat{H}_s + \hat{H}_{int}$, $\hat{H}_t$, $\hat{H}_s$ and
$\hat{H}_{int}$ referring to the tip, sample and their
interaction. $\hat{H}_s$ is obtained from the {\sc Fireball}-code
used to calculate the Si(112)$6\times1$-Ga surface; $\hat{H}_t$ is
calculated using the same DFT local-orbital code for a W-tip
having a pyramid with four atoms, attached to a W-(100) surface;
$\hat{H}_{int}$ is obtained using a dimer approximation, whereby
the different tip-sample hopping interactions, $\hat{T}_{ts}$, are
calculated from the dimer formed by the respective tip and sample
atoms whose interaction we want to obtain (it is shown in
Ref.~\onlinecite{Blanco:04} that this approximation yields a good
description of the STM images if orbitals with long-range tails
are used in the hopping calculations). A more detailed description
of our procedure to obtain theoretical STM images can be found in
Refs.~\onlinecite{Blanco:04, Blanco:05}. Making use of the total
Hamiltonian and the Keldysh Green-function techniques, we can
calculate the tunneling current from the following equation
\cite{Mingo:96}
\begin{eqnarray}
\label{cur} I \! \!  &=& \! \! \frac{4\pi e}{\hbar}  \! \! \!
\int_{-\infty}^{\infty} \! \! \! \! \! \! \!  d\omega \mbox{\rm
Tr} \bigl[ \hat T_{ts} \hat \rho_{ss}(\omega) \hat D_{ss}^r
(\omega) \hat T_{st} \hat \rho_{tt} (\omega) \hat D_{tt}^a(\omega)
\bigr]  \nonumber \\ & & \times (f_t(\omega) - f_s(\omega))
\end{eqnarray}

where

\begin{eqnarray}
\label{denom2} \hat D_{ss}^r = [\hat 1 - \hat T_{st} \hat
g_{tt}^r(\omega)\hat T_{ts} \hat g_{ss}^r(\omega) ]^{-1}
\end{eqnarray}
and
\begin{eqnarray}
\label{denom1} \hat D_{tt}^a = [\hat 1 - \hat T_{ts} \hat
g_{ss}^a(\omega)\hat T_{st} \hat g_{tt}^a(\omega) ]^{-1}
\end{eqnarray}

include all the interface multiple scattering processes. $Tr$
stands for the Trace of the current matrix. $\hat{g}_{ss}^{a(r)}$
and $\hat{g}_{tt}^{a(r)}$ are the advanced (retarded)
Green-functions of the sample and the tip, respectively
(calculated taking $ \hat{T}_{ts} = 0$); $\hat{\rho}_{ss}$ and
$\hat{\rho}_{tt}$ are the sample and tip density of states (also
for $\hat{T}_{ts} = 0$); and $f_t$ ($f_s$) the Fermi distribution
functions.

In the tunneling regime, $\hat{T}_{ts}$ is very small and
$\hat{D}_{ss}^{r}$ and $\hat{D}_{tt}^{a}$ can be replaced by
$\hat{I}$. In this limit, for zero-temperature, we recover the
following equation:

\begin{eqnarray}
I \! \!  &=& \! \! \frac{4\pi e}{\hbar}  \! \! \! \int_{E_F}^{E_F
+ eV} \! \! \! \! \! \! \!  d\omega \mbox{\rm Tr} \bigl[ \hat
T_{ts} \hat \rho_{ss}(\omega) \hat T_{st} \hat \rho_{tt} (\omega)
\bigr] \label{eqn:I}
\end{eqnarray}

which we have used to calculate the STM images of the different
surface structures.

We should comment that the detailed comparison between theory and
experimental results that we intend in this work requires the use
of  equation~(\ref{eqn:I}) instead of other simpler approaches
(like the Tersoff-Hamann formalism) that are common in the
literature. Our method includes a realistic description of the
geometry and the full electronic structure of the tip, and
incorporates quantitatively the influence of the tunneling
parameters (bias and current conditions) and the tip-sample
distance.\cite{Blanco:04,Blanco:05} This quantitative accuracy,
crucial to understand the contradictory experimental results in
terms of contrast and symmetry of the STM images of an apparently
simple system like O/Pd(111)-2$\times$2,\cite{Blanco:05} is
necessary in our case to discriminate among all the different
surface structures that have been analyzed in this work. Notice,
in particular, that we show below that our proposed model is fully
compatible with the STM images by Baski \emph{et al.}
\cite{Baski:99} provided that their tunneling parameters are used
in the simulation of the STM images. On top of these advantages,
we have to mention that our approach does not require a
significantly larger computational time than other simpler
methods, as equation~(\ref{eqn:I}) provides a very compact
procedure for calculating the tip-sample tunneling current that
takes full advantage of the LDOS ($\hat{\rho_{ss}}$ and
$\hat{\rho_{tt}}$) obtained from our DFT calculations.

\section{\label{sec:stmobs}STM observations}
\begin{figure}[ht]

  \centering
  \includegraphics[width=\columnwidth]{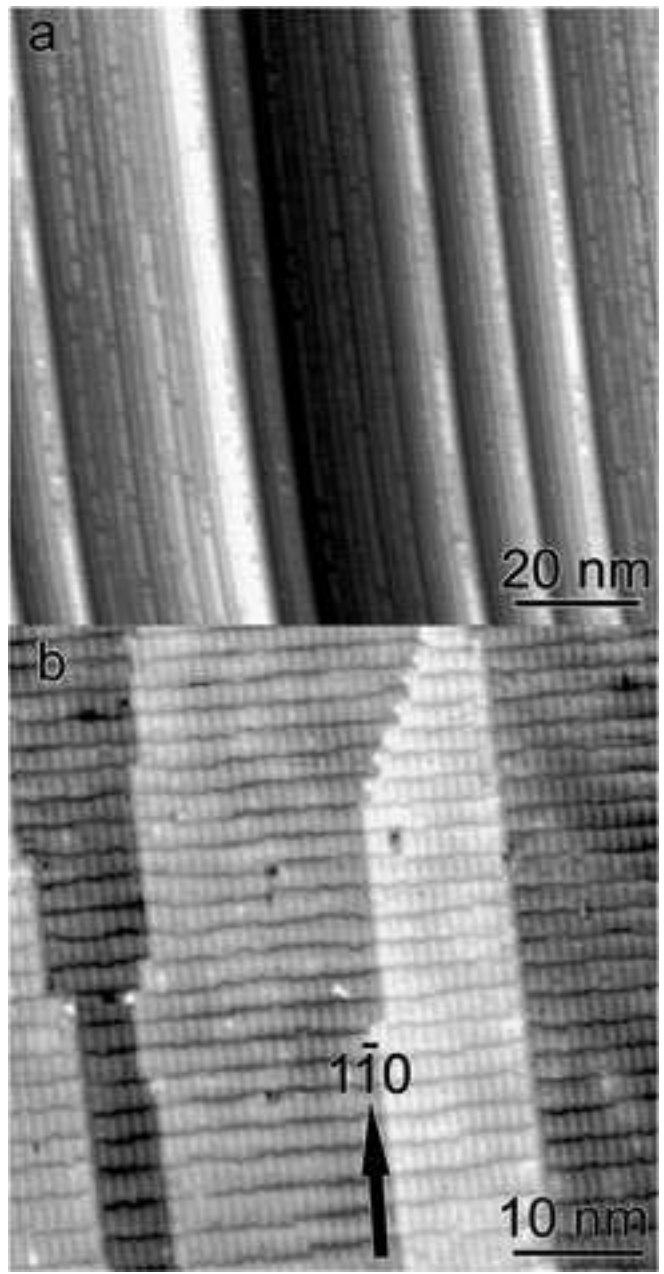}
  \caption{(\emph{a}) STM image of pristine Si(112). (\emph{b}) STM
  image of the Ga covered Si(112) surface. Tunneling conditions:
  1.5 V, 0.1 nA and 2 V, 0.1 nA, for (\emph{a}) and (\emph{b}), respectively.}
  \label{fig:STMlarge}
\end{figure}

The high index Si(112) surface is tilted 19.5$^\circ$ away from
the (111) surface towards (001). But the pristine Si(112) surface
is not thermodynamically stable, and breaks up into approximately
10 nm wide nano-facets of reconstructed (111)- and (337)-like
planes.\cite{Baski:95,Baski:96} An STM image of pristine Si(112)
is shown in Fig.~\ref{fig:STMlarge}(\emph{a}). One might expect
that metal deposition on this surface would result in the
formation of metallic nanowires in these prepatterned grooves.
However, it was shown by Baski \emph{et al.}\cite{Baski:96} that
upon deposition and post-annealing of a sub-monolayer amount of Ga
the faceted Si(112) surface undergoes a massive restructuring. It
returns to its basal (112) orientation, reconstructing as
described in the introduction. This preparation procedure of
deposition and postannealing of the surface resulted in a
reproducible self-limiting surface reconstruction with a
$6\times1$ unit cell.\cite{Jung:94} A large scale STM image of Ga
covered Si(112) is shown in Fig.~\ref{fig:STMlarge}(\emph{b}). The
nanoscale facets have developed into large anisotropic (112)
terraces that can extend for up to microns along
[1$\overline{1}$0] direction but are less than 100 nm wide. On the
terraces, the vacancy lines appear as dark trenches which run
perpendicular to the step edges present in this image. Closer
inspection reveals that these vacancy lines are not exactly
straight, but their position fluctuates around an average
position. As discussed below, this is due to the coexistence of
$6\times1$ and $5\times1$ units in the surface and due to the
presence of \emph{intrinsic} fluctuations in the vacancy
lines.\cite{Gonzalez:04}

\begin{figure}[ht]
  \centering
  \includegraphics[width=\columnwidth]{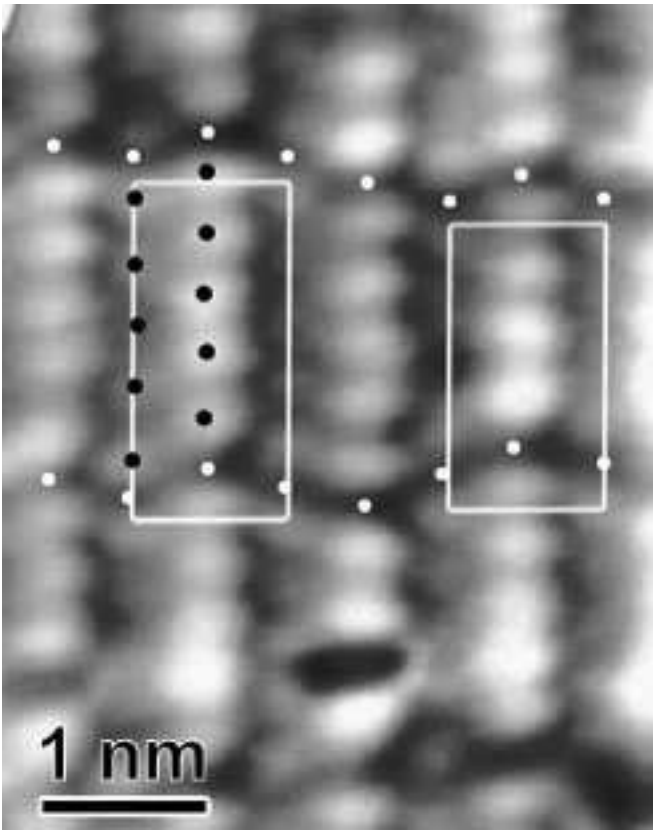}
  \caption{Empty state STM image of the Si(112)$n\times1$-Ga
surface. In this particular surface area both $5\times1$ and
$6\times1$ unit cells are present, as indicated. Atomic positions
are indicated with black dots. The position of the vacancies in
both atomic rows is indicated with white dots. Tunneling
conditions: 1.5 V, 0.2 nA.} \label{fig:SiGaHR}
\end{figure}

Detailed atomic resolution STM images were acquired to investigate
the atomic structure of the Si(112)$6\times1$-Ga surface. In
Fig.~\ref{fig:SiGaHR} an atomic resolution empty state STM image
is shown. Note that this particular area of the surface shows both
$5\times1$ and $6\times1$ unit cells as indicated in the figure.
Two parallel atom rows are observed per unit cell, running in the
[1$\overline{1}$0] direction, intersected by the quasi-periodical
vacancy lines. Comparing this image with the results of Baski
\emph{et al.},\cite{Baski:99} we observe the same spacing of the
brightest atom rows (\emph{i.e.} 9.4~\AA). Furthermore, the mixed
periodicities and the similar LEED pattern (see
Refs.~\onlinecite{Baski:99} and \onlinecite{Jung:94},
respectively), indicate that the same surface reconstruction is
studied here. Consequently we conclude that the brightest atom
rows in Fig.~\ref{fig:SiGaHR} are the same atom rows as imaged by
Baski \emph{et al}.\cite{Baski:99} (henceforth, the "step-edge Ga
row"). But in addition we observe a $2^{nd}$ atom row lying in
between the brighter rows. In terms of the JKP-model, this row of
atoms could be interpreted as the Si dangling bond row which might
form a quasi 1D band. However, these two parallel atomic lines
clearly form a \emph{zig-zag} pattern as indicated in
figures~\ref{fig:SiGaHR} (see also Fig.~\ref{fig:meander}), which
results in a structural \emph{asymmetry} in the vacancy line. This
is in contradiction with the JKP-model, which implies mirror plane
symmetry with respect to the (1$\overline{1}$0) plane in the
vacancy line.

\begin{figure}[ht]
  \centering
  \includegraphics[width=\columnwidth]{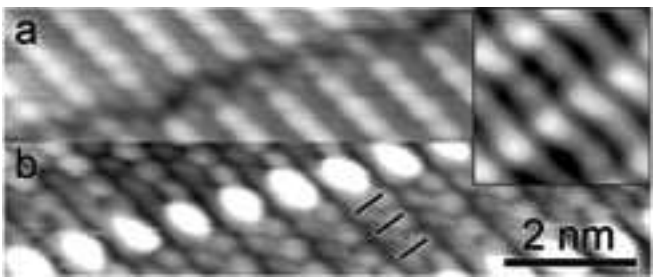}
  \caption{(\emph{a}) Empty state and (\emph{b}) filled state dual
  bias STM image of the Si(112)$6\times1$-Ga surface. In (\emph{b})
  the \emph{ladder} structure is indicated. Tunneling conditions:
  $\pm$ 1 V, 0.05 nA. The inset shows a filled state image with a
  slightly lower resolution. Tunneling conditions: -2 V, 0.1 nA.}
  \label{fig:SiGadualb}
\end{figure}

In Fig.~\ref{fig:SiGadualb} a set of registry aligned dual bias
images is presented. These images have been recorded
simultaneously on the same area of the surface, but with opposite
tunneling bias polarities resulting in a set of spatially
correlated empty and filled state images. In this case, the empty
state image, Fig.~\ref{fig:SiGadualb}(\emph{a}), has suffered from
a slight decrease in resolution, as compared with
Fig.~\ref{fig:SiGaHR}, but the asymmetry in the vacancy line is
still visible. In the filled state image,
Fig.~\ref{fig:SiGadualb}(\emph{b}), a relatively big, symmetric
protrusion prevents a detailed observation of the atomic structure
in the vacancy line. As in the empty state image, two parallel
atom rows are also visible in the filled state image. They form a
\emph{ladder} structure instead of the \emph{zig-zag} pattern of
the rows observed in the empty state image.

RBS measurements were performed to determine experimentally the
amount of Ga at the surface. Integration of the Ga peak in the
backscattered He spectrum yielded an amount of $9\pm1$ Ga atoms
per $6\times1$ unit cell, to be compared with 5 Ga per $6\times1$
unit cell for the JKP-model of Fig.~\ref{fig:structold}.

In summary, these experimental results (STM and RBS) consistently
show that the step-edge decorated JKP-model of the
Si(112)$6\times1$-Ga surface is at variance with the new
experimental observations. Consequently, the intuitive idea of
metal adatom step-edge decoration does not seem applicable for the
Ga/Si(112) interface.

\section{STM image simulations}
\label{sec:stmsim} %
Extensive DFT calculations were performed to identify the precise
atomic structure of the Si(112)$6\times1$-Ga surface. Using the
DFT local-orbital hamiltonian of the surface together with
non-equilibrium Keldysh Green function
techniques,\cite{Mingo:96,Jurczyszyn:01} we calculated theoretical
STM images of these most promising structures, which were then
compared with the high resolution experimental STM images. The
different Si(112)$6\times1$-Ga structures analyzed in this paper,
with Ga-coverages ranging from 5 to 11 Ga atoms per 6$\times$1
unit cell, have been generated starting from the JKP-model (see
Fig.~\ref{fig:structold}), in the following way: (a) replacing
some of the Si atoms in the Si-dangling-bond row by Ga atoms
(hereafter referred to as Ga terrace atoms); (b) replacing some of
the Ga atoms at the step edge by Si atoms; (c) considering also
the replacement of Si or Ga atoms on the step-edge and terrace
rows by vacancies and the addition of Ga or Si atoms in the
vacancy lines. In total, more than 40 new structures were fully
relaxed, their surface energies and electronic structures
calculated, and their corresponding theoretical STM images
obtained. In the following, we compare the theoretical STM images
of structures with the lowest total energies; a detailed chemical
potential analysis of the total energies is deferred to
Section~\ref{sec:competstruct}.

\begin{figure}[ht]
  \centering
\includegraphics[width=\columnwidth]{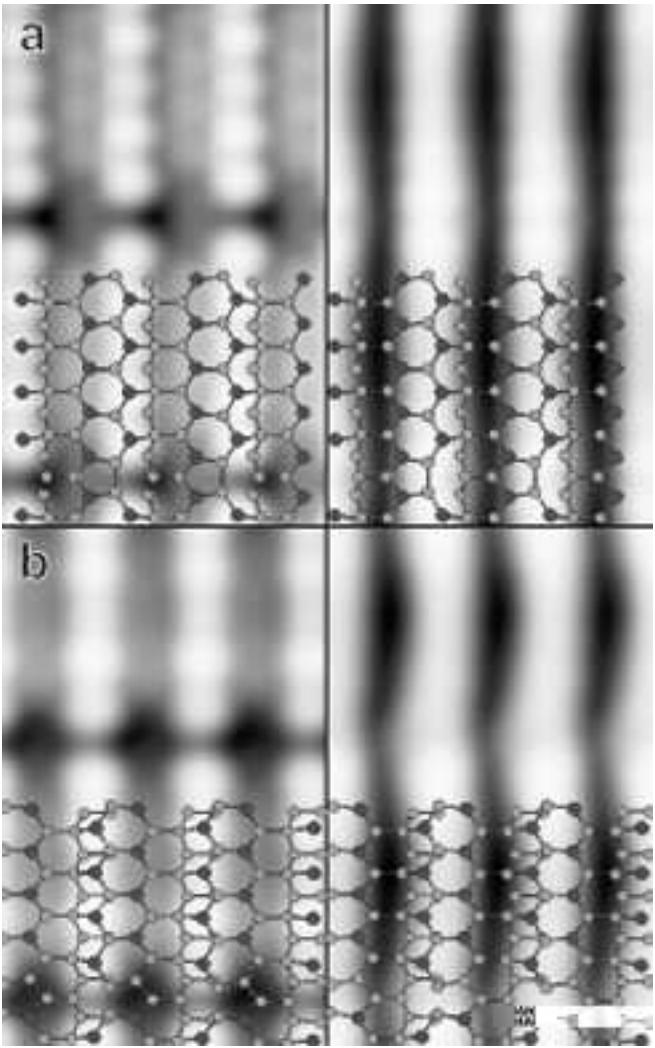}
  \caption{Simulated empty (left) and filled (right) state STM
  images of some of the structural models analyzed, with a ball
  and stick representation (top view) superimposed on top of the
  STM images. (\emph a) structural model with, per $6\times1$
  unit-cell, 6 Ga atoms in the terrace row and 5 Ga atoms in the
  step-edge row; (\emph b) a structural model with, per $6\times1$
  unit-cell, 5 Ga and 1 Si in the terrace row, and 5 Ga in the
  step-edge row. Ga atoms: dark, Si atoms: light.}
  \label{fig:other_models}
\end{figure}

Fig.~\ref{fig:other_models}(\emph{a}) and (\emph{b}) show two
examples of simulated STM images for some of these structural
models, with a top view of the corresponding atomic structure
superimposed. Fig.~\ref{fig:other_models}(\emph{a}) corresponds to
a structural model that contains, per $6\times1$ unit-cell, 6 Ga
atoms in the terrace row, and 5 Ga atoms plus a vacancy in the
step-edge row; in Fig.~\ref{fig:other_models}(\emph{b}) there are
5 Ga and one Si in the terrace row, and 5 Ga plus a vacancy in the
step-edge row. The simulated STM images for the different
structural models are compared in detail with the experimental
high-resolution STM images. For example,
Fig.~\ref{fig:other_models}(\emph{a}) (filled state) is similar to
the filled-state STM image obtained in Ref.~\onlinecite{Baski:99};
also the empty-state image of
Fig.~\ref{fig:other_models}(\emph{b}) is in good agreement with
the experimental STM image shown in Fig.~\ref{fig:SiGaHR}.
However, a detailed analysis of both empty and filled states
images, as well as registry aligned dual images (\emph{e.g.}
Fig.~\ref{fig:SiGadualb}), reveal that these models present some
inconsistency with the experimental high-resolution information.
For example, the structural model of
Fig.~\ref{fig:other_models}(\emph{a}) is symmetric with respect to
the vacancy line, in disagreement with Fig.~\ref{fig:SiGaHR}; in
the case of Fig.~\ref{fig:other_models}(\emph{b}) registry aligned
dual bias STM images show that the bright protrusion in the filled
state image is located in the vacancy line, aligned with the
brighter Ga row of the empty state image, while in the simulated
filled state-image it appears in between the two Ga rows.

\begin{figure}[ht]
  \centering
\includegraphics[width=\columnwidth]{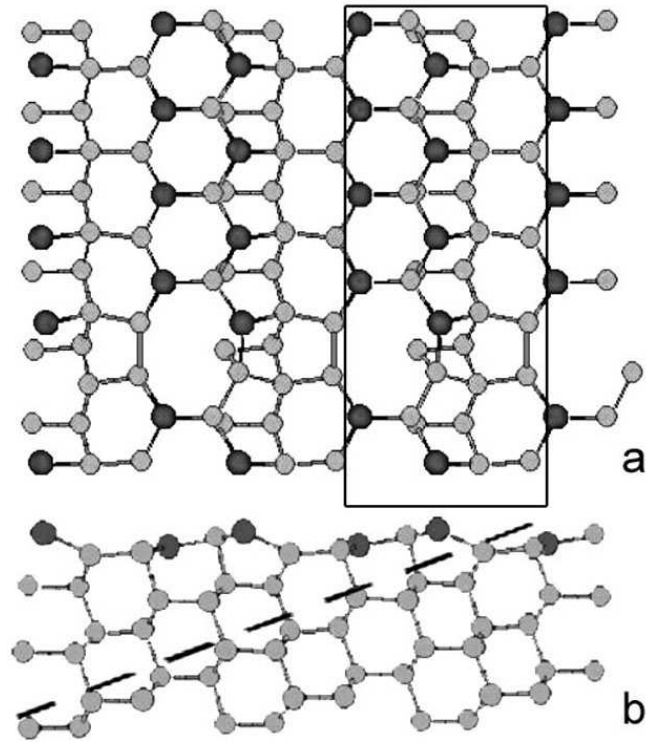}
  \caption{ Ball and stick representation of the energy minimized
  structure for the Si(112)$6\times1$-Ga surface; topview
  (\emph{a}), and sideview (\emph{b}). In (\emph{a}) a $6\times1$
  unit cell is indicated and in (\emph{b}) a (111) plane is
  indicated. Si atoms: light, Ga atoms: dark.}
  \label{fig:structmod}
\end{figure}

Thus, a detailed comparison of the theoretical STM images for the
different structural models with the experimental STM images was
performed. From this analysis, we concluded that the correct
atomic model for the Si(112)$6\times1$-Ga is the one shown in
Fig.~\ref{fig:structmod}. In this new structural model there are
10 Ga atoms per $6\times1$ unit-cell (to be compared with the RBS
determination of $9 \pm 1$ Ga atoms), forming two parallel rows,
in a \emph{zig-zag} configuration (see also
Fig.~\ref{fig:SiGaHR}). The upper row of step-edge Ga atoms
adsorbed at the (111)-like step is equivalent to the Ga row in the
JKP-model. But the Si dangling bond row in the JKP-model has been
replaced by a second row of Ga atoms (henceforth, the "terrace Ga
row"). Each Ga-row contains 5 Ga atoms per $6\times1$ unit-cell,
\emph{i.e.} there is a Ga-vacancy in each row. These vacancies are
placed at adjacent sites in the (\emph{zig-zag}) two rows, giving
rise to an asymmetry in the vacancy line, see
Figs.~\ref{fig:structmod} and \ref{fig:SiGaHR}.

Inside the vacancy lines, missing Ga atoms expose the underlying
Si atoms. These Si atoms rebond forming Si-Si dimers on the
terraces and Si-Ga dimers along the step edges in each unit cell.
Specifically, by rotating a step-edge Si atom toward the step-edge
Ga row, this Si atom can rebond to two neighboring Si atoms and a
step-edge Ga atom, forming a Si-Ga dimer with the latter. The
rebonding of the Si atoms in the vacancy line also implies that
the Ga vacancies on both Ga rows must be aligned. In contrast with
the \emph{tetravalent} Si atoms in the step-edge decorated
JKP-model, both the \emph{trivalent} Ga atoms in the three-fold
adsorption sites on the (111)-like terraces, and the exposed Si
atoms inside the vacancy lines present no unsaturated dangling
bonds; the resulting structure (Fig.~\ref{fig:structmod}) is fully
passivated and the surface is semiconducting.

\begin{figure}[ht]
  \centering
 \includegraphics[width=\columnwidth]{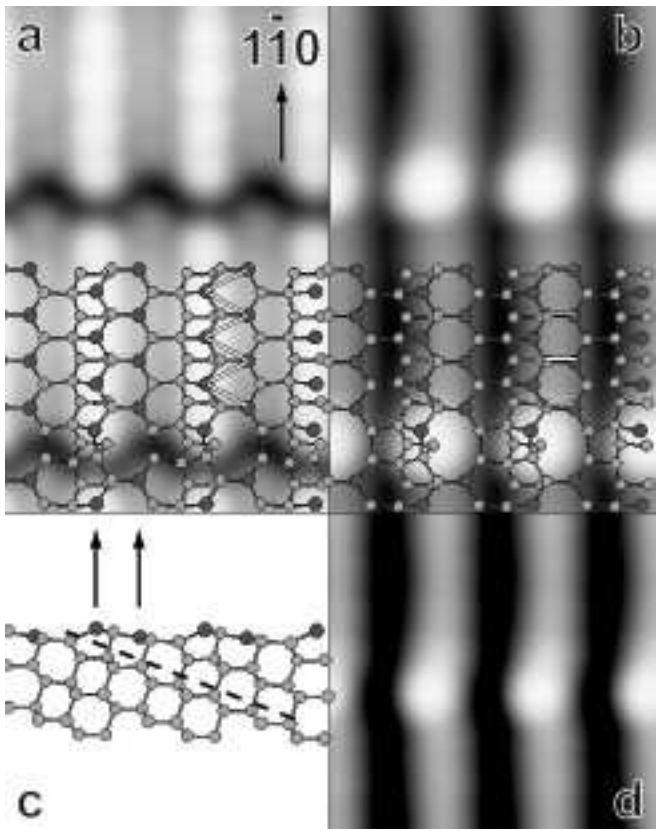}
  \caption{Simulated empty (\emph{a}) and filled (\emph{b}) state
  STM images of the \emph{zig-zag} model of
  Fig.~\ref{fig:structmod}. A top view of the ball and stick
  representation is superimposed on top of the STM images. Ga
  atoms: dark, Si atoms: light. \emph{Zig-zag} symmetry and
  \emph{ladder} symmetry indicated with white bars in (\emph{a})
  and (\emph{b}), respectively. Tunneling bias 2 V (\emph{a}), and
  - 1.3 V (\emph{b}). (\emph{c}) Side view of the proposed model.
  The (111) plane is indicated with a dotted line. (\emph{d})
  Simulated filled state image, -2 V.} \label{fig:simSTM}
\end{figure}

The calculated theoretical STM images corresponding to this
\emph{zig-zag} model are shown in Fig.~\ref{fig:simSTM}, with a
top view of the structural model superimposed on top. Both the
empty state and filled state images are in excellent agreement
with the experimental ones (see Fig.~\ref{fig:SiGaHR}). It shows
that the two atom rows imaged in the empty state STM images, are
indeed the step-edge Ga row and the terrace Ga row, ruling out the
formation of a Ga-atom step-edge decorated structure. In addition,
the asymmetry in the vacancy lines observed experimentally in the
empty state, is neatly reproduced in the simulated STM images. In
the simulated filled state image, fuzzy lines with a big,
symmetric protrusion inside the vacancy line are observed, in
agreement with the experimental images.
Fig.~\ref{fig:simSTM}({\emph b}) clearly shows that the big
protrusion corresponds to the Ga-Si dimer. Furthermore, it shows
that the fuzzy lines are originating from a Si-Ga bond on the
(111)-like terrace. They form a ladder configuration, in agreement
with the atomic resolution experimental image in
Fig.~\ref{fig:SiGadualb}. The only feature which was not
reproduced is the slightly higher apparent height, in the empty
state experimental images, of the Ga atoms in the two terrace Ga
rows directly adjacent to the vacancy line. Finally, we mention
that changing the tunneling conditions in the simulated STM images
(tip-sample distance, voltage), the experimental STM images of
Ref.~\onlinecite{Baski:99} can be recovered, as shown in
Fig.~\ref{fig:simSTM}({\emph d}).

\section{\label{sec:spectr} Spectroscopy}
\begin{figure}[ht]
  \centering
  \includegraphics[width=\columnwidth]{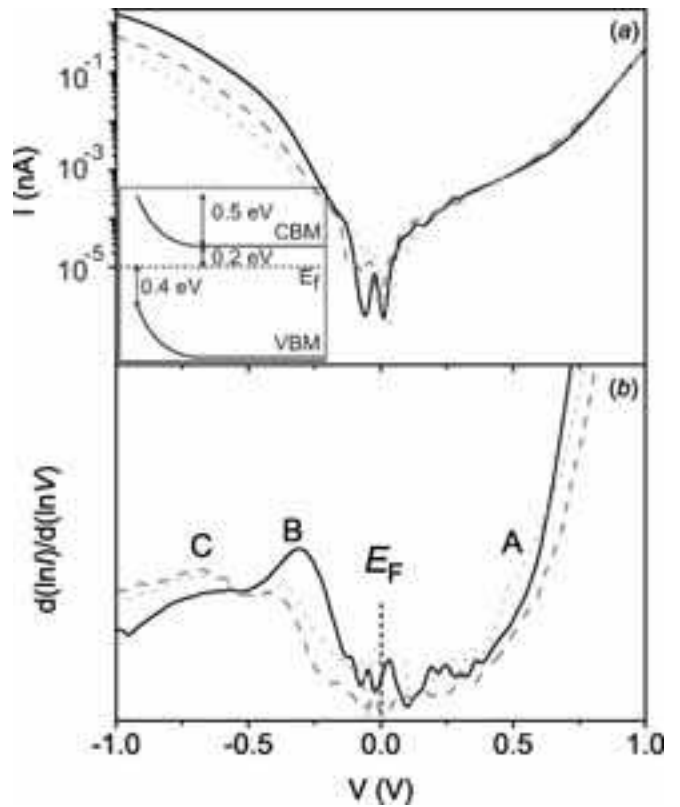}
  \caption{Semilogarithmic (\emph{a}) and normalized derivative
(\emph{b}) plots of $I-V$-curves averaged over the step-edge
(dotted) and terrace (dashed) atom rows, and the vacancy lines
(solid), respectively. STS setpoint: 1 V, 0.3 nA. The inset in
(\emph{a}) shows the bandstructure inferred from the data.}
  \label{fig:STS}
\end{figure}

\begin{figure}[ht]
  \centering
  \includegraphics[width=\columnwidth]{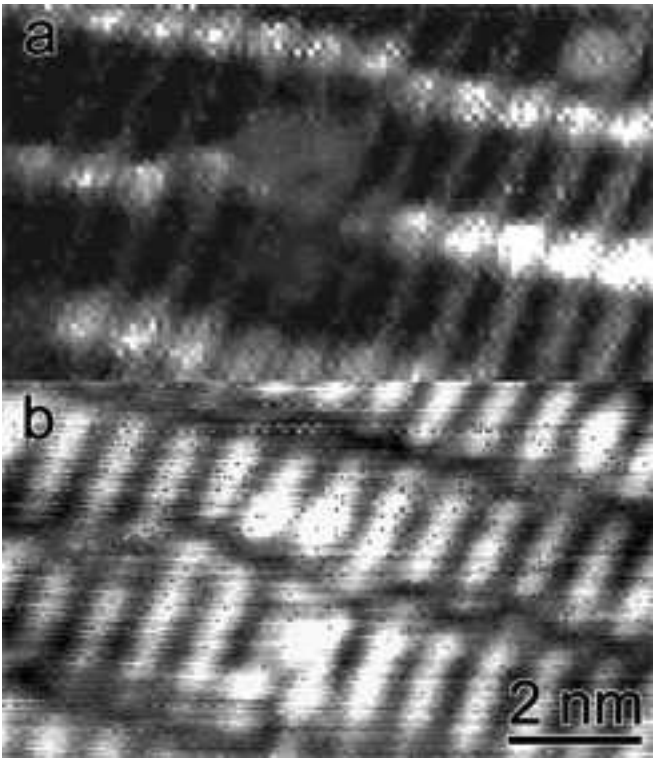}
  \caption{(\emph{a}) $\frac{\partial{I}}{\partial{V}}$-map of
  measured STS curves at -0.31 V. STS setpoint: 1 V, 0.3 nA.
  (\emph{b}) Corresponding topographic STM image. Tunneling
  conditions 1 V, 0.3 nA.}
  \label{fig:didv}
\end{figure}

We also have studied this surface reconstruction with scanning
tunneling spectroscopy. While imaging the surface with constant
tunneling current, at every third data point an $I-V$ curve is
measured with the feedback loop switched off during this $I-V$
measurement. In Fig.~\ref{fig:STS} we have averaged $I-V$ curves
measured on the upper atom rows, the lower atom rows, and on the
vacancy lines separately (the respective areas being determined
from the empty state STM image). This results in three curves,
representing the electronic structure on the terrace and step-edge
Ga rows, and the electronic structure inside the vacancy line. At
bias voltages below the bulk conduction band minimum, the
tunneling current is limited by thermionic emission, as is evident
from the linear increase of the $log(I)-V$ curve,\cite{Sze:81} up
to the conduction band minimum (CBM) at 0.7 V, see
Fig.~\ref{fig:STS}(\emph{a}). Consequently, the bulk valence band
maximum (VBM) at the surface should be located at $\sim-0.4$ V,
implying an upward band bending of $\sim0.5$ eV for this n-type
specimen ($10^{15} cm^{-3}$), $E_{f} - E_{VBM}$ and $E_{CBM} -
E_{f}$ being $\sim0.4$ and $\sim0.7$ eV at the surface,
respectively, as shown in the inset of
Fig.~\ref{fig:STS}(\emph{a}). These data are consistent with the
measured surface photovoltage in
Refs.~\onlinecite{Yoo:02,unpub:04}. In
Fig.~\ref{fig:STS}(\emph{b}) we have plotted the normalized
derivative of the three $I-V$ curves (\emph{i.e.}
$\frac{\partial{\ln{I}}}{\partial{\ln{V}}}$), originating from the
three different areas within the unit cell. These tunneling
spectra are proportional to the local density of states (LDOS) at
the respective areas of the sample surface over which the
averaging took place (see
Refs.~\onlinecite{Feenstra:86,Tersoff:83,Tersoff:85}). In all
curves, there is no DOS at the Fermi level, but a gap of exists
between the filled and empty state bands showing that indeed the
surface is semiconducting. The tunneling spectra on the two Ga
atom rows appears to be similar in shape, whereas the tunneling
spectra inside the vacancy line deviates from the former two. The
leading edge of the total tunneling spectrum in the filled state
spectrum arises from a state (B) that is located mainly inside the
vacancy lines at about $\sim-0.3$ eV. The two Ga rows feature a
broad filled state at higher binding energy (C). In the empty
state tunneling spectra the two atom rows show a small shoulder
(A) at about $\sim0.6$ eV, just below the bulk CBM. Thus a surface
band gap of $\sim0.9$ eV is deduced. These experimental data are
entirely consistent with the presence of two equivalent, threefold
coordinated rows of Ga atoms at these positions, fully passivating
the surface. Notice that we inferred the surface band gap from the
separation between peaks A and B using the peak position or
centroids, and \emph{not} the onsets. The justification for this
procedure comes from a detailed comparison with theoretical STS
data, as will be discussed below.

In addition, we have constructed a
$\frac{\partial{I}}{\partial{V}}$-map of the STS measurements. In
a $\frac{\partial{I}}{\partial{V}}$-map, the value of the
derivative of the $I-V$-curves at a certain voltage $V$ is plotted
as a two-dimensional image, with the $x$ and $y$ coordinates
corresponding to the topographic STM image. In
Fig.~\ref{fig:didv}(\emph{a}) the derivative of the $I-V$-curves
at -0.31 V is plotted, the corresponding empty state STM image is
shown in Fig.~\ref{fig:didv}(\emph{b}). Indeed, the largest slope
in the $I-V$-curves at -0.31 V is located inside the vacancy lines
(\emph{i.e.} here the largest increase in tunneling current is
observed, corresponding to the largest LDOS as compared to the
LDOS at this specific energy at other locations on the surface).
Only a very small intensity variation is observed perpendicular to
the atom rows, consistent with the similar filled state tunneling
spectra on the two Ga rows in Fig.~\ref{fig:STS}(\emph{a}).

\begin{figure}[ht]
  \centering
 \includegraphics[width=\columnwidth]{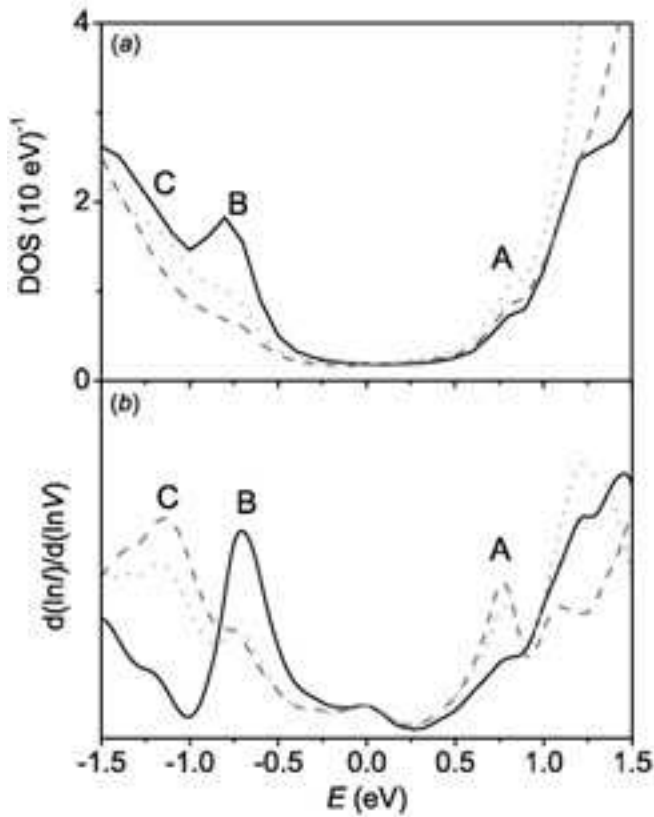}
  \caption{(\emph{a}) Calculated LDOS ({\sc Fireball96}), averaged
over the step-edge (dotted), terrace (dashed) Ga-rows, and vacancy
line (solid), respectively. The LDOS at the vacancy line is
calculated as the average density of states of the Ga-Si dimer and
Si-Si dimer. (\emph{b}) Normalized derivative of the $I-V$-curves
shown in (\emph{a}). A broadening of 0.1 eV has been applied. Note
the larger energy scale as compared to Fig.~\ref{fig:STS} due to
the overestimation of the gap in the local orbital calculation.}
  \label{fig:LDOS}
\end{figure}

From the structural model and its spatially resolved DOS,
theoretical STS curves were calculated, see
Fig.~\ref{fig:LDOS}(\emph{b}). As for the STM images, these
results are obtained using the LO-DFT {\sc Fireball96} hamiltonian
of the surface and the Keldysh Green function approach. The
corresponding calculated LDOS of the \emph{zig-zag} model,
averaged over different areas is shown in
Fig.~\ref{fig:LDOS}(\emph{a}) (a broadening of 0.1 eV has been
used). The calculated STS curves shown in
Fig.~\ref{fig:LDOS}(\emph{b}) were obtained by placing the tip
over the respective areas, calculating the current as a function
of a voltage sweep using the calculated LDOS
(Fig.~\ref{fig:LDOS}), and averaging over the areas of interest.
The {\sc Fireball96} local orbital calculations employ a minimal
basis set, resulting typically in band gaps that are too large.
Nonetheless, excellent qualitative agreement exists between the
calculated STS curves and normalized derivatives of the
experimental STS curves (Fig.~\ref{fig:STS}(\emph{b})). The
calculated LDOS and STS curves confirm that the large peak B just
beneath the band gap indeed is mainly associated with states that
are located on the Si-Ga dimer inside the vacancy lines. The two
Ga rows contribute almost equally to a broad peak in the DOS at
higher binding energy (C) and a small shoulder in the DOS just
above the band gap (A), in full agreement with the normalized
derivative of the tunneling spectra in
Fig.~\ref{fig:STS}(\emph{b}). The empty state DOS is very similar
for both Ga rows. Consequently the $\sim0.4$~\AA higher appearance
of the Ga atoms at the step edge in the empty state image is due
to their on average higher atomic positions, and thus the empty
state STM image reflects the real surface topography at these
voltages.

\begin{figure}[ht]
  \centering
  \includegraphics[width=\columnwidth]{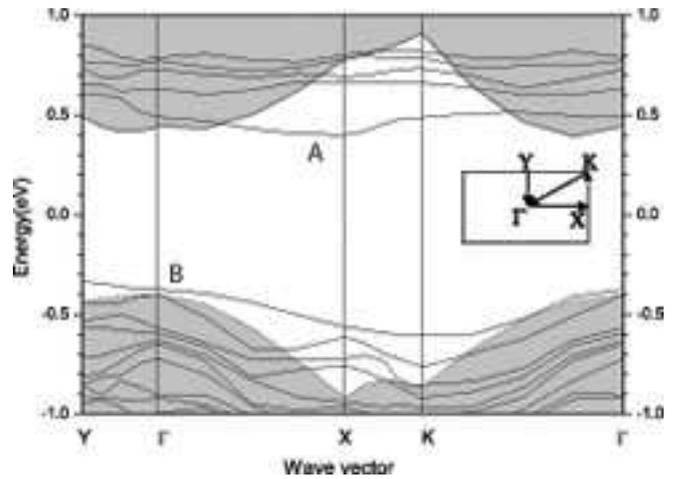}
  \caption{Calculated band structure ({\sc Castep}) of the
\emph{zig-zag} structural model. Surface states labelled A and B
are mentioned in the text. The shaded area shows the bulk
projected bandstructure of the Si substrate. The inset shows the
surface Brillouin zone probed.}
  \label{fig:bandstructure}
\end{figure}

We have calculated the surface bandstructure using both the LO and
PW-DFT methods, within the LDA for exchange-correlation
contributions. While the LO calculation overestimates the value of
the bulk band gap, the PW calculation typically underestimates the
band gap. In the LO bandstructure (not shown) a surface state band
gap of 1.2 eV is obtained between the states A and B, while the
separation between peaks A and B in the calculated STS is close to
1.4 eV. This suggests that in order to measure the band gap, it is
reasonably accurate to use the peak positions instead of the
(poorly defined) peak onsets in the experimental
$\frac{\partial{\ln{I}}}{\partial{\ln{V}}}$ curves. The observed
splitting between A and B in the experimental curves is $\sim0.9$
eV indicating that the experimental band gap is $\sim0.8$ eV.
Fig.~\ref{fig:bandstructure} shows the bandstructure as calculated
with the PW-DFT code, showing a surface band gap of 0.77 eV
between states A and B.\footnote{Note that the bulk projected gap
for the PW (LDA) calculation is not equal to $\sim0.67$ eV, as a
fully converged PW (LDA) calculation should give. This is caused
by the fact that the bands are plotted along symmetry directions
in the surface Brillouin zone that do not contain the k-point
corresponding to the bulk Si CBM.} The calculations place filled
state B slightly above the VBM, in agreement with the experimental
observation. On the other hand, the empty surface state A is
located at or slightly above the CBM according to the PW-DFT
calculations, while experimentally state A appears slightly below
the bulk CBM; see Fig~\ref{fig:STS}(\emph{b}). The PW gap of 0.77
eV is comparable to the experimental band gap of 0.8 eV. However,
the precise location of state A in the calculations directly
affects the value of the band gap.

Despite the fact that the Ga/Si(112) overlayer appears to be quasi
two-dimensional in \emph{atomic} structure, the \emph{electronic}
structure of this overlayer is quasi one-dimensional. The Ga-atom
induced surface band A disperses around the $X$-point minimum in
the upper part of the band gap. The dispersion of this band near
the $X$-point yields an effective mass of $m^{*} \sim1.48~m_{e}$
along the $X - \Gamma$ direction,  and $m^{*} \sim0.15~m_{e}$
along $X - K$. This indicates a quasi-one-dimensional dispersion.
This Ga-band is initially empty but could perhaps be populated in
a controllable way, using a biased gate electrode, or a heavily
\emph{n}-type doped substrate, making the Si(112)$6\times1$-Ga
surface a promising system for the experimental study of electron
transport in one dimensional atomic wires.

\section{\label{sec:competstruct} Competing structures}
The discussion above shows that, to elucidate the precise atomic
structure of a complex surface like Si(112)$6\times1$-Ga from the
comparison of theoretical and experimental STM images, it is
necessary to use high-resolution experimental STM images,
including registry aligned dual bias information and STS data,
combined with state of the art theoretical STM simulations. These
simulations were performed on the subset of possible structures
that were deemed most realistic on the basis of total energy
considerations. In this section, we explore the relative stability
of the various structures, which gives a more physical basis to
the proposed structural model.

In general, the precise stoichiometry of the surface is not known,
and thus the analysis of the relative stability of different
structural models requires the calculation of the surface energy
$F$ as a function of the different chemical potentials.

\subsection{\label{sec:chempot} Chemical potential analysis} For
the analysis of the relative stabilities of the various
structures, we need to calculate the surface energy $F = E_{tot} -
\mu_{Ga} N_{Ga} -\mu_{Si} N_{Si}$, where $E_{tot}$ is the total
energy per unit-cell, $\mu_{Ga}$, $\mu_{Si}$ are the Ga and Si
chemical potentials, and $N_{Ga}$, $N_{Si}$ are the number of Ga
and Si atoms in the unit-cell. For $\mu_{Si}$ we use the total
energy (per atom) of bulk-Si (\emph{i.e.} the surface is in
equilibrium with the substrate). The value of $\mu_{Ga}$ is not
determined by the substrate, but it can be estimated analyzing the
experimental conditions (see below).

Fig.~\ref{fig:F_mu} shows the surface energy $F = E_{tot} -
\mu_{Ga} N_{Ga} -\mu_{Si} N_{Si}$ as a function of $\mu_{Ga}$. In
this figure we use the structural model of
Fig.~\ref{fig:structmod} as reference, and plot $F$ for some of
the most promising models, as calculated with the PW code ({\sc
Castep}). In order to estimate the value of $\mu_{Ga}$ we have to
analyze the experimental conditions of the Ga deposition. In the
"one-step" process the ($6 \times 1$)-phase is formed under a Ga
flux from the effusion cell with the sample held at a temperature
of $T = 825$ K. At this temperature the incoming flux of Ga atoms
is balanced by a flux of Ga atoms desorbing from the surface, thus
establishing a quasi-equilibrium. This allows us to relate the
chemical potential $\mu_{Ga}$ in the effusion cell with the
chemical potential $\mu_{Ga}$ at the sample. The chemical
potential in the effusion cell may be approximated by the total
energy of bulk-Ga, $\mu_{Ga}(bulk)$, \emph{i.e.} the Ga vapor in
the effusion cell is in equilibrium with the solid.\footnote{The
chemical potential of liquid Ga is approximated by the chemical
potential of solid, bulk Ga at T=0K. Finite temperature
corrections which include the enthalpy of melting and the
integrated heat capacity are very small ($<0.1$ eV) and have been
neglected.}

Considering also the equilibrium between the sample and the Ga
vapor in contact with the sample, we can estimate the chemical
potential at the sample

\[
\mu_{Ga} = \mu_{Ga}(bulk) - k_B T ln (\frac{p_c}{p_s})
\]
where $p_c$ is the Ga vapor pressure in the effusion cell and
$p_s$ the Ga vapor pressure at the sample. Since the effusion cell
flux is proportional to its vapour pressure, $p_c$, times the cell
aperture area, and the sample flux is also proportional to its
corresponding vapour pressure, $p_s$, times the sample area, we
conclude that $(p_c/p_s) \sim10^{2} (10^{3})$, and $\mu_{Ga} =
\mu_{Ga-bulk} - 0.32 (0.48)$ eV. In Fig.~\ref{fig:F_mu} we see
that for this range of $\mu_{Ga}$ the structural model of
Fig.~\ref{fig:structmod} presents the lowest surface energy $F$.
This result strongly supports our conclusion that the structural
model for the Si(112)$6\times1$-Ga surface is the one depicted in
Fig.~\ref{fig:structmod}.

\begin{figure}[ht]
  \centering
  \includegraphics[width=\columnwidth]{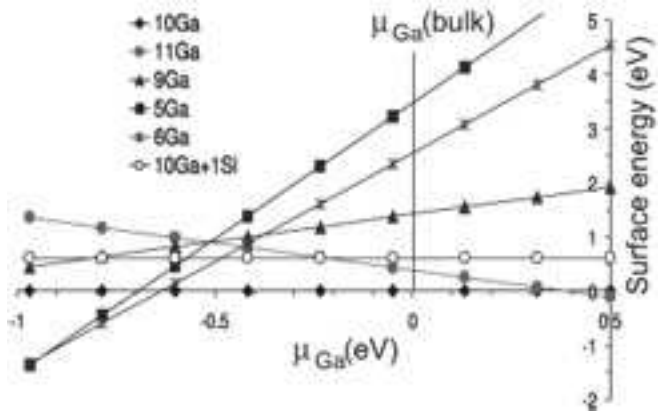}
  \caption{Surface energy as a function of the Ga chemical potential ({\sc Castep}).
  The Ga chemical potential is plotted relative to the chemical
  potential in bulk Ga. 5Ga (filled squares) is the step-edge
  decorated JKP-model} \label{fig:F_mu}
\end{figure}

Comparing our model with the step-edge decorated JKP-model, an
important difference is that the new structural model
(Fig.~\ref{fig:structmod}) presents no partially-filled dangling
bonds as discussed above. The stability of the new model is
related to the \emph{full} passivation of the substrate, removing
all dangling bonds, and the associated decrease in surface free
energy. The results shown in Fig.~\ref{fig:F_mu} suggest, however,
that the JKP-model might be stabilized for very low $\mu_{Ga}$
values. We should stress that our theoretical analysis has been
directed to search for surface atomic structures that could
explain the experimental results (STM and RBS) for the
Si(112)$6\times1$-Ga surface, and thus surface structures with
lower Ga coverages, that should be favored for low $\mu_{Ga}$
values have not been analyzed as thoroughly as those with
coverages close to 9-10 Ga atoms/$6\times1$ unit cell.
Nevertheless, we may perform a simple analysis, comparing the
surface energy of the step-edge decorated JKP-model, with the
surface energy of a simple hypothetical surface: half the surface
is covered with the structure of Fig.~\ref{fig:structmod}
(\emph{i.e.} both step-edge and terrace Ga rows) while the other
half consists of clean Si(112). Both the step-edge decorated JKP
system and this hypothetical half-half case present the same Ga
coverage and thus the same behavior of $F$ as a function of
$\mu_{Ga}$ (\emph{i.e.} the same slope in Fig.~\ref{fig:F_mu}).
This comparison reveals that the hypothetical case is lower in
energy (by $\sim$ 0.7 eV/(10 Ga atoms), for all $\mu_{Ga}$ values,
showing that the step-edge decorated case is unlikely to be
stabilized at lower Ga coverage, and phase separation into bare
Si(112) and the Si(112)$6\times1$-Ga \emph{zig-zag} surface will
occur instead. Note that the facetting of the unstable Si(112)
surface, which was not accounted for in this calculation, would
increase this energy difference, making phase separation even more
favorable compared to the step-edge decorated JKP-model.

Another possible scenario for obtaining a step-edge decorated Ga
row would be to use the experimentally observed fully passivated
Si(112)$6\times1$-Ga surface (Fig.~\ref{fig:structmod}) as
starting point, and try to \emph{kinetically} stabilize a
\emph{metastable} step-edge decorated structure by selectively
desorbing the Ga atoms from the terraces. Experimentally, this
might happen in the 'two-step' preparation procedure, see
Section~\ref{sec:exp}. We have studied this possibility by
calculating desorption energies of terrace and step-edge Ga atoms
from the Si(112)$6\times1$-Ga surface (Fig.~\ref{fig:structmod}).
In particular, we have considered removing the Ga atoms close to
the vacancy line as well as the replacement of those Ga atoms by
Si atoms. In both cases the desorption energies are lower by
$\sim0.7$ eV for step-edge Ga atoms than for Ga atoms on the
terraces. This result suggests that a metastable step-edge
decorated structure likely can not be achieved by thermally
desorbing the Ga terrace atoms.

\subsection{\label{sec:fluct} Intrinsic structural disorder}

\begin{figure}[ht]
  \centering
  \includegraphics[width=\columnwidth]{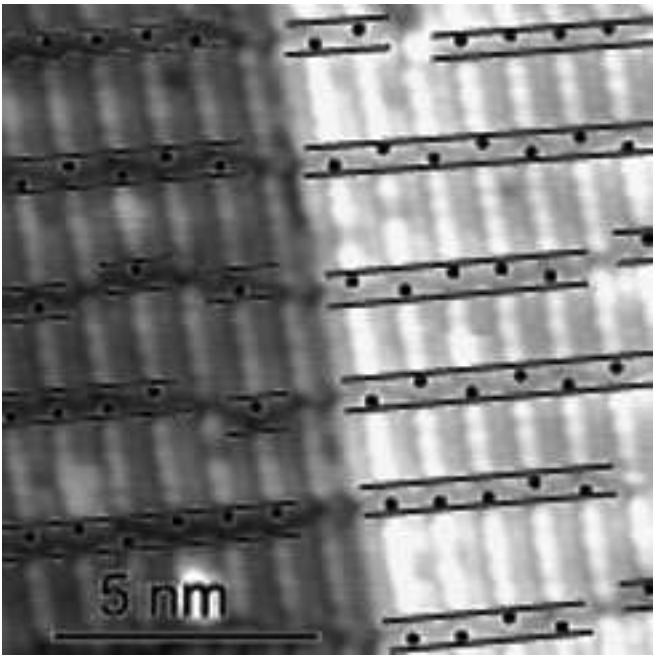}
  \caption{Detailed STM image, showing the two contributions to
  the meandering of the vacancy lines. Straight lines are drawn
  through domains with unit cells of the same size. Occasional
  jumps in the lines are due to different unit cell sizes, as
  explained in the text. Dots are placed on the Ga atom of the
  Ga-Si dimer at the vacancy of the step-edge rows, showing the
  fluctuations due to the intrinsic disorder of the random orientation of the Ga-Si
  dimer.} \label{fig:meander}
\end{figure}

\begin{figure}[ht]
  \centering
  \includegraphics[width=\columnwidth]{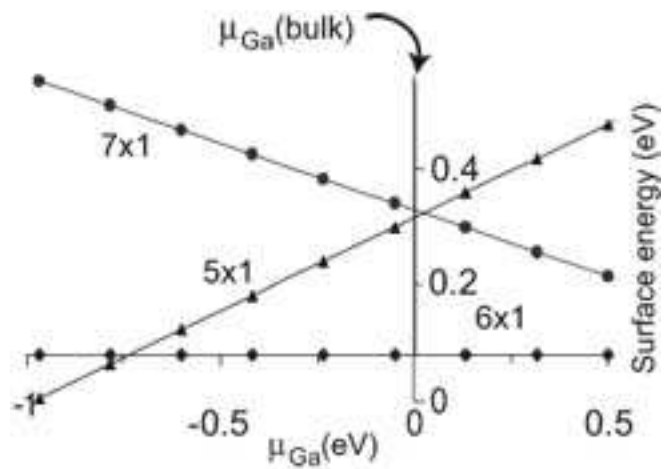}
  \caption{Free energy as a function of the Ga chemical potential for
5x1, 6x1 and 7x1. The Ga chemical potential is plotted relative to
the chemical potential in bulk Ga. Note the difference in scale on
the abcissa as compared to Fig.~\ref{fig:F_mu}.}
  \label{fig:F_5x1}
\end{figure}

In the experimental STM images, the vacancy lines are not exactly
straight, but some meandering is observed, as shown in
Fig.~\ref{fig:meander} (see also
Figs.~\ref{fig:STMlarge}(\emph{b}) and Fig.~\ref{fig:SiGaHR}). The
new structural model is able to fully explain the experimentally
observed meandering.\cite{Gonzalez:04} It was proposed by Erwin
\emph{et al.}\cite{Erwin:99} that this meandering of the vacancy
lines could be explained by the co-existence of $6\times1$ and
$5\times1$ unit cells on the surface. This results in occasional
steps in the vacancy lines, equivalent to the observed meandering
of the dimer-vacancy lines on the Ge covered Si(001)
surfaces.\cite{Chen:94} We have analyzed the stability of the
\emph{zig-zag} structural model as a function of the longitudinal
periodicity. Fig.~\ref{fig:F_5x1} shows the surface energies $F$
of this model for different periodicities: $5\times1$, $6\times1$
and $7\times1$. The $5\times1$ surface corresponds to 4 Ga atoms
in each Ga-row between vacancy lines, while the $7\times1$ surface
presents 6 Ga atoms in each row between vacancy lines. For our
estimated range of $\mu_{Ga}$ the $6\times1$ surface presents the
lowest $F$, while the $5\times1$ is only 0.1-0.2 eV higher, per
$6\times1$ unit-cell. This small energy difference should lead to
the experimental observation of $5\times1$ unit cells. Indeed
these unit cells are frequently observed in the experimental
images, see Fig.~\ref{fig:meander}. It thus appears that the
predictions from the 1D Frenkel-Kontorova model regarding the
vacancy-line spacing in the step-edge decorated
structure\cite{Erwin:99} also apply to the quasi 2D \emph{zig-zag}
arrangement of Ga atoms presented here. This conclusion is not
very surprising as it was concluded in Ref.~\onlinecite{Erwin:99}
that the strain induced by the size difference of Ga and Si
completely dominates the energetics of the periodicities.
Apparently this conclusion still holds when an extra Ga row is
added.

However, careful investigation of the experimental images reveals
that the meandering of the vacancy lines as observed in
Fig.~\ref{fig:STMlarge} is not only due to the competing
longitudinal periodicities. Instead, for large sections of the
surface, the \emph{terrace Ga rows} are perfectly periodic in the
$\times1$ direction with (n-1) Ga atoms per terrace Ga row in the
n$\times1$ unit cell. But in these ordered sections the number of
Ga atoms in the \emph{step-edge Ga rows} appears to fluctuate
between n-2 and n. The proposed structural model perfectly
explains these \emph{intrinsic} fluctuations (\emph{i.e.}
fluctuations within a n$\times1$ domain); they are related to the
orientation of the Si-Ga dimer in the step-edge rows. The twofold
symmetry of the 112 substrate in the [1$\overline{1}$0] direction
is broken by the Si-Ga dimers, resulting in two degenerate
orientations of these dimers. The energy associated with
interchanging the atoms of a Si-Ga dimer has been
calculated\cite{Gonzalez:04} to be less than 10 meV per
$12\times1$ unit cell. This small energy difference explains the
appearance of frequent meandering in the aligned vacancies in the
step-edge Ga rows, thus accounting for the majority of the
fluctuations in the vacancy lines observed in the experimental
images. The absence of these fluctuations in the filled state
images (compare Figs.~\ref{fig:SiGadualb}(\emph{a}) and (\emph{b})
and also Figs.~\ref{fig:didv}(\emph{a}) and (\emph{b})) is the
result of the fact that the bright protrusion in the vacancy line
appears in the center of the Ga-Si dimer, making its appearance
insensitive to the orientation of the Ga-Si dimer.

\section{\label{sec:concl}Summary and conclusions}
The $6\times1$ reconstruction of Ga on vicinal Si(112) was studied
with STM, STS, RBS, and extensive DFT calculations. High
resolution STM experiments revealed an asymmetry in the vacancy
lines of the Si(112)$6\times1$-Ga surface that is inconsistent
with the JKP-model of step-edge decoration. STS measurements also
rule out formation of quasi 1D metal wires while RBS experiments
indicated a Ga coverage twice as large as previously inferred from
the JKP model. Extensive DFT calculations were used to analyze the
relative stability of more than forty structures, taking the
chemical potential of the Ga adsorbate into account. Theoretical
STM images were calculated for the most promising structures and
compared in detail with the experimental STM images.

A new structure emerged containing 10 Ga atoms per $6\times1$ unit
cell. The Ga atoms decorate the step edge and passivate the
terrace atoms, thereby forming a \emph{zig-zag} pattern. Excellent
agreement between experimental and theoretical STM and STS data
confirmed the validity of the proposed \emph{zig-zag} model and
demonstrate the power of such a comparison. Ga atoms are threefold
coordinated and Si dangling bonds are all passivated so the
surface is semiconducting. The "broken bond orbitals" inside the
vacancy lines rebond to form Si-Ga and Si-Si dimers. The observed
meandering of the vacancy lines originates from thermal
fluctuations between the two symmetry-degenerate orientations of
the Si-Ga dimer, in conjunction with thermal fluctuations between
competing $6\times1$ and $5\times1$ units.

While step-edge decoration of vicinal metal surfaces
works,\cite{Gambardella:02} the observed drive toward chemical
passivation suggests that step-edge decoration of vicinal
semiconductors is not a viable method to produce 1D metal wires.
Although this general conclusion remains to be tested further, it
is clear that partially-filled dangling bonds on the terraces of
vicinal surfaces are always greatly reduced in number or
eliminated altogether in the reconstruction.

As shown in this paper, predictive calculations along these lines
should always take into account the chemical potential of the
adsorbate. The latter depends on the experimental preparation
conditions (see e.g. equation~(\ref{eqn:I})). Successful
prediction of systems with perfect 1D metal adatom step-edge
decoration could facilitate the quest for the experimental
realization of Luttinger liquids in such systems, possibly
enabling a convincing proof of spin-charge separation with angle
resolved photoemission spectroscopy.

\section{\label{sec:level1}Acknowledgements}
This work is part of the research programme of the 'Stichting voor
Fundamenteel Onderzoek der Materie (FOM)', which is financially
supported by the 'Nederlandse Organisatie voor Wetenschappelijk
Onderzoek (NWO)'. This work was sponsored in part by the NSF under
contract No. DMR-0244570, the Ministerio de Ciencia y
Tecnolog\'{i}a (Spain) under grants No. MAT2001-0665 and
MAT2004-01271. One of us, S.R., wishes to acknowledge the Royal
Netherlands Academy of Arts and Sciences. We thank T.M. Klapwijk
for his stimulating support, and the AMOLF institute in Amsterdam
for performing the RBS experiments. Oak Ridge National Laboratory
is managed by UT-Battelle, LLC, for the US Department of Energy
under contract No. DE-AC-05-00OR22725.


\begin{thebibliography}{39}
\expandafter\ifx\csname
natexlab\endcsname\relax\def\natexlab#1{#1}\fi
\expandafter\ifx\csname bibnamefont\endcsname\relax
  \def\bibnamefont#1{#1}\fi
\expandafter\ifx\csname bibfnamefont\endcsname\relax
  \def\bibfnamefont#1{#1}\fi
\expandafter\ifx\csname citenamefont\endcsname\relax
  \def\citenamefont#1{#1}\fi
\expandafter\ifx\csname url\endcsname\relax
  \def\url#1{\texttt{#1}}\fi
\expandafter\ifx\csname
urlprefix\endcsname\relax\def\urlprefix{URL }\fi
\providecommand{\bibinfo}[2]{#2}
\providecommand{\eprint}[2][]{\url{#2}}

\bibitem[{\citenamefont{Postma et~al.}(2000)\citenamefont{Postma, de~Jonge,
  Yao, and Dekker}}]{Dekker:00}
\bibinfo{author}{\bibfnamefont{H.~W.~C.} \bibnamefont{Postma}},
  \bibinfo{author}{\bibfnamefont{M.}~\bibnamefont{de~Jonge}},
  \bibinfo{author}{\bibfnamefont{Z.}~\bibnamefont{Yao}}, \bibnamefont{and}
  \bibinfo{author}{\bibfnamefont{C.}~\bibnamefont{Dekker}},
  \bibinfo{journal}{Phys. Rev. B} \textbf{\bibinfo{volume}{62}},
  \bibinfo{pages}{R10653} (\bibinfo{year}{2000}).

\bibitem[{\citenamefont{Grioni et~al.}(1999)\citenamefont{Grioni, Vobornik,
  Zwick, and Margaritondo}}]{Grioni:99}
\bibinfo{author}{\bibfnamefont{M.}~\bibnamefont{Grioni}},
  \bibinfo{author}{\bibfnamefont{I.}~\bibnamefont{Vobornik}},
  \bibinfo{author}{\bibfnamefont{F.}~\bibnamefont{Zwick}}, \bibnamefont{and}
  \bibinfo{author}{\bibfnamefont{G.}~\bibnamefont{Margaritondo}},
  \bibinfo{journal}{J. Electron Spectrosc. Relat. Phenom.}
  \textbf{\bibinfo{volume}{100}}, \bibinfo{pages}{313} (\bibinfo{year}{1999}).

\bibitem[{\citenamefont{Voit}(1994)}]{Voit:94}
\bibinfo{author}{\bibfnamefont{J.}~\bibnamefont{Voit}}, \bibinfo{journal}{Rep.
  Prog. Phys.} \textbf{\bibinfo{volume}{58}}, \bibinfo{pages}{977}
  (\bibinfo{year}{1994}).

\bibitem[{\citenamefont{Himpsel et~al.}(2001)\citenamefont{Himpsel, Altmann,
  Bennewitz, Crain, Kirakosian, Lin, and McChesney}}]{Himpsel:01}
\bibinfo{author}{\bibfnamefont{F.~J.} \bibnamefont{Himpsel}},
  \bibinfo{author}{\bibfnamefont{K.~N.} \bibnamefont{Altmann}},
  \bibinfo{author}{\bibfnamefont{R.}~\bibnamefont{Bennewitz}},
  \bibinfo{author}{\bibfnamefont{J.~N.} \bibnamefont{Crain}},
  \bibinfo{author}{\bibfnamefont{A.}~\bibnamefont{Kirakosian}},
  \bibinfo{author}{\bibfnamefont{J.-L.} \bibnamefont{Lin}}, \bibnamefont{and}
  \bibinfo{author}{\bibfnamefont{J.~L.} \bibnamefont{McChesney}},
  \bibinfo{journal}{J. Phys.: Condens. Matter} \textbf{\bibinfo{volume}{13}},
  \bibinfo{pages}{11097} (\bibinfo{year}{2001}).

\bibitem[{\citenamefont{Crain et~al.}(2004)\citenamefont{Crain, McChesney,
  Zheng, Gallagher, Snijders, Bissen, Gundelach, Erwin, and
  Himpsel}}]{Himpsel:04}
\bibinfo{author}{\bibfnamefont{J.~N.} \bibnamefont{Crain}},
  \bibinfo{author}{\bibfnamefont{J.~L.} \bibnamefont{McChesney}},
  \bibinfo{author}{\bibfnamefont{F.}~\bibnamefont{Zheng}},
  \bibinfo{author}{\bibfnamefont{M.~C.} \bibnamefont{Gallagher}},
  \bibinfo{author}{\bibfnamefont{P.~C.} \bibnamefont{Snijders}},
  \bibinfo{author}{\bibfnamefont{M.}~\bibnamefont{Bissen}},
  \bibinfo{author}{\bibfnamefont{C.}~\bibnamefont{Gundelach}},
  \bibinfo{author}{\bibfnamefont{S.~C.} \bibnamefont{Erwin}}, \bibnamefont{and}
  \bibinfo{author}{\bibfnamefont{F.~J.} \bibnamefont{Himpsel}},
  \bibinfo{journal}{Phys. Rev. B} \textbf{\bibinfo{volume}{69}},
  \bibinfo{pages}{125401} (\bibinfo{year}{2004}).

\bibitem[{\citenamefont{Starowicz et~al.}(2002)\citenamefont{Starowicz, Gallus,
  Pillo, and Baer}}]{Baer:02}
\bibinfo{author}{\bibfnamefont{P.}~\bibnamefont{Starowicz}},
  \bibinfo{author}{\bibfnamefont{O.}~\bibnamefont{Gallus}},
  \bibinfo{author}{\bibfnamefont{T.}~\bibnamefont{Pillo}}, \bibnamefont{and}
  \bibinfo{author}{\bibfnamefont{Y.}~\bibnamefont{Baer}},
  \bibinfo{journal}{Phys. Rev. Lett.} \textbf{\bibinfo{volume}{89}},
  \bibinfo{pages}{256402} (\bibinfo{year}{2002}).

\bibitem[{\citenamefont{Segovia et~al.}(1999)\citenamefont{Segovia, Purdie,
  Hengsberger, and Baer}}]{Baer:99}
\bibinfo{author}{\bibfnamefont{P.}~\bibnamefont{Segovia}},
  \bibinfo{author}{\bibfnamefont{D.}~\bibnamefont{Purdie}},
  \bibinfo{author}{\bibfnamefont{M.}~\bibnamefont{Hengsberger}},
  \bibnamefont{and} \bibinfo{author}{\bibfnamefont{Y.}~\bibnamefont{Baer}},
  \bibinfo{journal}{Nature} \textbf{\bibinfo{volume}{402}},
  \bibinfo{pages}{504} (\bibinfo{year}{1999}).

\bibitem[{\citenamefont{Yeom et~al.}(1999)\citenamefont{Yeom, Takeda,
  Rotenberg, Matsuda, Horikoshi, Schaefer, Lee, Kevan, Ohta, Nagao
  et~al.}}]{Yeom:99}
\bibinfo{author}{\bibfnamefont{H.~W.} \bibnamefont{Yeom}},
  \bibinfo{author}{\bibfnamefont{S.}~\bibnamefont{Takeda}},
  \bibinfo{author}{\bibfnamefont{E.}~\bibnamefont{Rotenberg}},
  \bibinfo{author}{\bibfnamefont{I.}~\bibnamefont{Matsuda}},
  \bibinfo{author}{\bibfnamefont{K.}~\bibnamefont{Horikoshi}},
  \bibinfo{author}{\bibfnamefont{J.}~\bibnamefont{Schaefer}},
  \bibinfo{author}{\bibfnamefont{C.~M.} \bibnamefont{Lee}},
  \bibinfo{author}{\bibfnamefont{S.~D.} \bibnamefont{Kevan}},
  \bibinfo{author}{\bibfnamefont{T.}~\bibnamefont{Ohta}},
  \bibinfo{author}{\bibfnamefont{T.}~\bibnamefont{Nagao}},
  \bibnamefont{et~al.}, \bibinfo{journal}{Phys. Rev. Lett.}
  \textbf{\bibinfo{volume}{82}}, \bibinfo{pages}{4898} (\bibinfo{year}{1999}).

\bibitem[{\citenamefont{Ahn et~al.}(2004)\citenamefont{Ahn, Yeom, Cho, and
  Park}}]{Ahn:04}
\bibinfo{author}{\bibfnamefont{J.~R.} \bibnamefont{Ahn}},
  \bibinfo{author}{\bibfnamefont{H.~W.} \bibnamefont{Yeom}},
  \bibinfo{author}{\bibfnamefont{E.~S.} \bibnamefont{Cho}}, \bibnamefont{and}
  \bibinfo{author}{\bibfnamefont{C.~Y.} \bibnamefont{Park}},
  \bibinfo{journal}{Phys. Rev. B} \textbf{\bibinfo{volume}{69}},
  \bibinfo{pages}{233311} (\bibinfo{year}{2004}).

\bibitem[{\citenamefont{Bunk et~al.}(1999)\citenamefont{Bunk, Falkenberg,
  Zeysing, Lottermoser, Johnson, Nielsen, Berg-Rasmussen, Baker, and
  Feidenhans'l}}]{Bunk:99}
\bibinfo{author}{\bibfnamefont{O.}~\bibnamefont{Bunk}},
  \bibinfo{author}{\bibfnamefont{G.}~\bibnamefont{Falkenberg}},
  \bibinfo{author}{\bibfnamefont{J.~H.} \bibnamefont{Zeysing}},
  \bibinfo{author}{\bibfnamefont{L.}~\bibnamefont{Lottermoser}},
  \bibinfo{author}{\bibfnamefont{R.~L.} \bibnamefont{Johnson}},
  \bibinfo{author}{\bibfnamefont{M.}~\bibnamefont{Nielsen}},
  \bibinfo{author}{\bibfnamefont{F.}~\bibnamefont{Berg-Rasmussen}},
  \bibinfo{author}{\bibfnamefont{J.}~\bibnamefont{Baker}}, \bibnamefont{and}
  \bibinfo{author}{\bibfnamefont{R.}~\bibnamefont{Feidenhans'l}},
  \bibinfo{journal}{Phys. Rev. B} \textbf{\bibinfo{volume}{59}},
  \bibinfo{pages}{12228} (\bibinfo{year}{1999}).

\bibitem[{\citenamefont{Robinson et~al.}(2002)\citenamefont{Robinson, Bennett,
  and Himpsel}}]{Himpsel:02}
\bibinfo{author}{\bibfnamefont{I.~K.} \bibnamefont{Robinson}},
  \bibinfo{author}{\bibfnamefont{P.~A.} \bibnamefont{Bennett}},
  \bibnamefont{and} \bibinfo{author}{\bibfnamefont{F.~J.}
  \bibnamefont{Himpsel}}, \bibinfo{journal}{Phys. Rev. Lett.}
  \textbf{\bibinfo{volume}{88}}, \bibinfo{pages}{096104}
  (\bibinfo{year}{2002}).

\bibitem[{\citenamefont{Gonz\'{a}lez et~al.}(2004)\citenamefont{Gonz\'{a}lez,
  Snijders, Ortega, P\'{e}rez, Flores, Rogge, and Weitering}}]{Gonzalez:04}
\bibinfo{author}{\bibfnamefont{C.}~\bibnamefont{Gonz\'{a}lez}},
  \bibinfo{author}{\bibfnamefont{P.~C.} \bibnamefont{Snijders}},
  \bibinfo{author}{\bibfnamefont{J.}~\bibnamefont{Ortega}},
  \bibinfo{author}{\bibfnamefont{R.}~\bibnamefont{P\'{e}rez}},
  \bibinfo{author}{\bibfnamefont{F.}~\bibnamefont{Flores}},
  \bibinfo{author}{\bibfnamefont{S.}~\bibnamefont{Rogge}}, \bibnamefont{and}
  \bibinfo{author}{\bibfnamefont{H.~H.} \bibnamefont{Weitering}},
  \bibinfo{journal}{Phys. Rev. Lett.} \textbf{\bibinfo{volume}{93}},
  \bibinfo{pages}{126106} (\bibinfo{year}{2004}).

\bibitem[{\citenamefont{Crain et~al.}(2003)\citenamefont{Crain, Kirakosian,
  Altmann, Bromberger, Erwin, McChesney, Lin, and Himpsel}}]{Crain:03}
\bibinfo{author}{\bibfnamefont{J.~N.} \bibnamefont{Crain}},
  \bibinfo{author}{\bibfnamefont{A.}~\bibnamefont{Kirakosian}},
  \bibinfo{author}{\bibfnamefont{K.~N.} \bibnamefont{Altmann}},
  \bibinfo{author}{\bibfnamefont{C.}~\bibnamefont{Bromberger}},
  \bibinfo{author}{\bibfnamefont{S.~C.} \bibnamefont{Erwin}},
  \bibinfo{author}{\bibfnamefont{J.~L.} \bibnamefont{McChesney}},
  \bibinfo{author}{\bibfnamefont{J.-L.} \bibnamefont{Lin}}, \bibnamefont{and}
  \bibinfo{author}{\bibfnamefont{F.~J.} \bibnamefont{Himpsel}},
  \bibinfo{journal}{Phys. Rev. Lett.} \textbf{\bibinfo{volume}{90}},
  \bibinfo{pages}{176805} (\bibinfo{year}{2003}).

\bibitem[{\citenamefont{Ahn et~al.}(2003)\citenamefont{Ahn, Yeom, Yoon, and
  Lyo}}]{Ahn:03}
\bibinfo{author}{\bibfnamefont{J.~R.} \bibnamefont{Ahn}},
  \bibinfo{author}{\bibfnamefont{H.~W.} \bibnamefont{Yeom}},
  \bibinfo{author}{\bibfnamefont{H.~S.} \bibnamefont{Yoon}}, \bibnamefont{and}
  \bibinfo{author}{\bibfnamefont{I.-W.} \bibnamefont{Lyo}},
  \bibinfo{journal}{Phys. Rev. Lett.} \textbf{\bibinfo{volume}{91}},
  \bibinfo{pages}{196403} (\bibinfo{year}{2003}).

\bibitem[{\citenamefont{Matsuda et~al.}(2003)\citenamefont{Matsuda,
  Hengsberger, Baumberger, Greber, Yeom, and Osterwalder}}]{Matsuda:03}
\bibinfo{author}{\bibfnamefont{I.}~\bibnamefont{Matsuda}},
  \bibinfo{author}{\bibfnamefont{M.}~\bibnamefont{Hengsberger}},
  \bibinfo{author}{\bibfnamefont{F.}~\bibnamefont{Baumberger}},
  \bibinfo{author}{\bibfnamefont{T.}~\bibnamefont{Greber}},
  \bibinfo{author}{\bibfnamefont{H.~W.} \bibnamefont{Yeom}}, \bibnamefont{and}
  \bibinfo{author}{\bibfnamefont{J.}~\bibnamefont{Osterwalder}},
  \bibinfo{journal}{Phys. Rev. B} \textbf{\bibinfo{volume}{68}},
  \bibinfo{pages}{195319} (\bibinfo{year}{2003}).

\bibitem[{\citenamefont{Kanagawa et~al.}(2003)\citenamefont{Kanagawa, Hobara,
  Matsuda, Tanikawa, Natori, and Hasegawa}}]{Hasegawa:03}
\bibinfo{author}{\bibfnamefont{T.}~\bibnamefont{Kanagawa}},
  \bibinfo{author}{\bibfnamefont{R.}~\bibnamefont{Hobara}},
  \bibinfo{author}{\bibfnamefont{I.}~\bibnamefont{Matsuda}},
  \bibinfo{author}{\bibfnamefont{T.}~\bibnamefont{Tanikawa}},
  \bibinfo{author}{\bibfnamefont{A.}~\bibnamefont{Natori}}, \bibnamefont{and}
  \bibinfo{author}{\bibfnamefont{S.}~\bibnamefont{Hasegawa}},
  \bibinfo{journal}{Phys. Rev. Lett.} \textbf{\bibinfo{volume}{91}},
  \bibinfo{pages}{036805} (\bibinfo{year}{2003}).

\bibitem[{\citenamefont{Jung et~al.}(1993)\citenamefont{Jung, Kaplan, and
  Prokes}}]{Jung:93}
\bibinfo{author}{\bibfnamefont{T.~M.} \bibnamefont{Jung}},
  \bibinfo{author}{\bibfnamefont{R.}~\bibnamefont{Kaplan}}, \bibnamefont{and}
  \bibinfo{author}{\bibfnamefont{S.~M.} \bibnamefont{Prokes}},
  \bibinfo{journal}{Surf. Sci.} \textbf{\bibinfo{volume}{289}},
  \bibinfo{pages}{L577} (\bibinfo{year}{1993}).

\bibitem[{\citenamefont{Jung et~al.}(1994)\citenamefont{Jung, Prokes, and
  Kaplan}}]{Jung:94}
\bibinfo{author}{\bibfnamefont{T.~M.} \bibnamefont{Jung}},
  \bibinfo{author}{\bibfnamefont{S.~M.} \bibnamefont{Prokes}},
  \bibnamefont{and} \bibinfo{author}{\bibfnamefont{R.}~\bibnamefont{Kaplan}},
  \bibinfo{journal}{J. Vac. Sci. Technol. A} \textbf{\bibinfo{volume}{12}},
  \bibinfo{pages}{1838} (\bibinfo{year}{1994}).

\bibitem[{\citenamefont{Yater et~al.}(1995)\citenamefont{Yater, Shih, and
  Idzerda}}]{Yater:95}
\bibinfo{author}{\bibfnamefont{J.~E.} \bibnamefont{Yater}},
  \bibinfo{author}{\bibfnamefont{A.}~\bibnamefont{Shih}}, \bibnamefont{and}
  \bibinfo{author}{\bibfnamefont{Y.~U.} \bibnamefont{Idzerda}},
  \bibinfo{journal}{Phys. Rev. B} \textbf{\bibinfo{volume}{51}},
  \bibinfo{pages}{R7365} (\bibinfo{year}{1995}).

\bibitem[{\citenamefont{Baski et~al.}(1999)\citenamefont{Baski, Erwin, and
  Whitman}}]{Baski:99}
\bibinfo{author}{\bibfnamefont{A.~A.} \bibnamefont{Baski}},
  \bibinfo{author}{\bibfnamefont{S.~C.} \bibnamefont{Erwin}}, \bibnamefont{and}
  \bibinfo{author}{\bibfnamefont{L.~J.} \bibnamefont{Whitman}},
  \bibinfo{journal}{Surf. Sci.} \textbf{\bibinfo{volume}{423}},
  \bibinfo{pages}{L265} (\bibinfo{year}{1999}).

\bibitem[{\citenamefont{Erwin et~al.}(1999)\citenamefont{Erwin, Baski, Whitman,
  and Rudd}}]{Erwin:99}
\bibinfo{author}{\bibfnamefont{S.~C.} \bibnamefont{Erwin}},
  \bibinfo{author}{\bibfnamefont{A.~A.} \bibnamefont{Baski}},
  \bibinfo{author}{\bibfnamefont{L.~J.} \bibnamefont{Whitman}},
  \bibnamefont{and} \bibinfo{author}{\bibfnamefont{R.~E.} \bibnamefont{Rudd}},
  \bibinfo{journal}{Phys. Rev. Lett.} \textbf{\bibinfo{volume}{83}},
  \bibinfo{pages}{1818} (\bibinfo{year}{1999}).

\bibitem[{\citenamefont{Yoo et~al.}(2002)\citenamefont{Yoo, Tang, Sprunger,
  Benito, Ortega, Flores, Snijders, Demeter, and Weitering}}]{Yoo:02}
\bibinfo{author}{\bibfnamefont{K.}~\bibnamefont{Yoo}},
  \bibinfo{author}{\bibfnamefont{S.~J.} \bibnamefont{Tang}},
  \bibinfo{author}{\bibfnamefont{P.~T.} \bibnamefont{Sprunger}},
  \bibinfo{author}{\bibfnamefont{I.}~\bibnamefont{Benito}},
  \bibinfo{author}{\bibfnamefont{J.}~\bibnamefont{Ortega}},
  \bibinfo{author}{\bibfnamefont{F.}~\bibnamefont{Flores}},
  \bibinfo{author}{\bibfnamefont{P.~C.} \bibnamefont{Snijders}},
  \bibinfo{author}{\bibfnamefont{M.~C.} \bibnamefont{Demeter}},
  \bibnamefont{and} \bibinfo{author}{\bibfnamefont{H.~H.}
  \bibnamefont{Weitering}}, \bibinfo{journal}{Surf. Sci.}
  \textbf{\bibinfo{volume}{514}}, \bibinfo{pages}{100} (\bibinfo{year}{2002}).

\bibitem[{\citenamefont{Baski and Whitman}(1996)}]{Baski:96}
\bibinfo{author}{\bibfnamefont{A.~A.} \bibnamefont{Baski}} \bibnamefont{and}
  \bibinfo{author}{\bibfnamefont{L.~J.} \bibnamefont{Whitman}},
  \bibinfo{journal}{J. Vac. Sci. Technol. B} \textbf{\bibinfo{volume}{14}},
  \bibinfo{pages}{992} (\bibinfo{year}{1996}).

\bibitem[{\citenamefont{Glembocki and Proke}(1997)}]{Glembocki:97}
\bibinfo{author}{\bibfnamefont{O.~J.} \bibnamefont{Glembocki}}
  \bibnamefont{and} \bibinfo{author}{\bibfnamefont{S.~M.} \bibnamefont{Proke}},
  \bibinfo{journal}{Appl. Phys. Lett.} \textbf{\bibinfo{volume}{71}},
  \bibinfo{pages}{2355} (\bibinfo{year}{1997}).

\bibitem[{\citenamefont{Demkov et~al.}(1995)\citenamefont{Demkov, Ortega,
  Sankey, and Grumbach}}]{Demkov:95}
\bibinfo{author}{\bibfnamefont{A.~A.} \bibnamefont{Demkov}},
  \bibinfo{author}{\bibfnamefont{J.}~\bibnamefont{Ortega}},
  \bibinfo{author}{\bibfnamefont{O.~F.}~\bibnamefont{Sankey}}, \bibnamefont{and}
  \bibinfo{author}{\bibfnamefont{M.~P.}~\bibnamefont{Grumbach}},
  \bibinfo{journal}{Phys. Rev. B} \textbf{\bibinfo{volume}{52}},
  \bibinfo{pages}{1618} (\bibinfo{year}{1995}).

\bibitem[{\citenamefont{Sankey and Niklewski}(1989)}]{Sankey:89}
\bibinfo{author}{\bibfnamefont{O.~F.} \bibnamefont{Sankey}} \bibnamefont{and}
  \bibinfo{author}{\bibfnamefont{D.~J.} \bibnamefont{Niklewski}},
  \bibinfo{journal}{Phys. Rev. B} \textbf{\bibinfo{volume}{40}},
  \bibinfo{pages}{3979} (\bibinfo{year}{1989}).

\bibitem[{CAS()}]{CASTEP}
\bibinfo{note}{CASTEP 4.2 Academic version, licensed under the UKCP-MSI
  Agreement, 1999}.

\bibitem[{\citenamefont{Mingo et~al.}(1996)\citenamefont{Mingo, Jurczyszyn,
  Garcia-Vidal, Saiz-Pardo, de~Andres, Flores, Wu, and More}}]{Mingo:96}
\bibinfo{author}{\bibfnamefont{N.}~\bibnamefont{Mingo}},
  \bibinfo{author}{\bibfnamefont{L.}~\bibnamefont{Jurczyszyn}},
  \bibinfo{author}{\bibfnamefont{F.~J.} \bibnamefont{Garcia-Vidal}},
  \bibinfo{author}{\bibfnamefont{R.}~\bibnamefont{Saiz-Pardo}},
  \bibinfo{author}{\bibfnamefont{P.~L.} \bibnamefont{de~Andres}},
  \bibinfo{author}{\bibfnamefont{F.}~\bibnamefont{Flores}},
  \bibinfo{author}{\bibfnamefont{S.~Y.} \bibnamefont{Wu}}, \bibnamefont{and}
  \bibinfo{author}{\bibfnamefont{W.}~\bibnamefont{More}},
  \bibinfo{journal}{Phys. Rev. B} \textbf{\bibinfo{volume}{54}},
  \bibinfo{pages}{2225} (\bibinfo{year}{1996}).

\bibitem[{\citenamefont{Jurczyszyn et~al.}(2001)\citenamefont{Jurczyszyn,
  J.Ortega, P\'{e}rez, and Flores}}]{Jurczyszyn:01}
\bibinfo{author}{\bibfnamefont{L.}~\bibnamefont{Jurczyszyn}},
  \bibinfo{author}{\bibnamefont{J.Ortega}},
  \bibinfo{author}{\bibfnamefont{R.}~\bibnamefont{P\'{e}rez}},
  \bibnamefont{and} \bibinfo{author}{\bibfnamefont{F.}~\bibnamefont{Flores}},
  \bibinfo{journal}{Surf. Sci.} \textbf{\bibinfo{volume}{482-5}},
  \bibinfo{pages}{1350} (\bibinfo{year}{2001}).

\bibitem[{\citenamefont{Blanco et~al.}(2004)\citenamefont{Blanco, Gonz\'{a}lez,
  Jel\'{i}nek, Ortega, Flores, and P\'{e}rez}}]{Blanco:04}
\bibinfo{author}{\bibfnamefont{J.~M.} \bibnamefont{Blanco}},
  \bibinfo{author}{\bibfnamefont{C.}~\bibnamefont{Gonz\'{a}lez}},
  \bibinfo{author}{\bibfnamefont{P.}~\bibnamefont{Jel\'{i}nek}},
  \bibinfo{author}{\bibfnamefont{J.}~\bibnamefont{Ortega}},
  \bibinfo{author}{\bibfnamefont{F.}~\bibnamefont{Flores}}, \bibnamefont{and}
  \bibinfo{author}{\bibfnamefont{R.}~\bibnamefont{P\'{e}rez}},
  \bibinfo{journal}{Phys. Rev. B} \textbf{\bibinfo{volume}{70}},
  \bibinfo{pages}{085405} (\bibinfo{year}{2004}).

\bibitem[{\citenamefont{Blanco et~al.}(2005)\citenamefont{Blanco, Gonz\'{a}lez,
  Jel\'{i}nek, Ortega, Flores, P\'{e}rez, Rose, Salmeron, M\'{e}ndez,
  Wintterlin et~al.}}]{Blanco:05}
\bibinfo{author}{\bibfnamefont{J.~M.} \bibnamefont{Blanco}},
  \bibinfo{author}{\bibfnamefont{C.}~\bibnamefont{Gonz\'{a}lez}},
  \bibinfo{author}{\bibfnamefont{P.}~\bibnamefont{Jel\'{i}nek}},
  \bibinfo{author}{\bibfnamefont{J.}~\bibnamefont{Ortega}},
  \bibinfo{author}{\bibfnamefont{F.}~\bibnamefont{Flores}},
  \bibinfo{author}{\bibfnamefont{R.}~\bibnamefont{P\'{e}rez}},
  \bibinfo{author}{\bibfnamefont{M.}~\bibnamefont{Rose}},
  \bibinfo{author}{\bibfnamefont{M.}~\bibnamefont{Salmeron}},
  \bibinfo{author}{\bibfnamefont{J.}~\bibnamefont{M\'{e}ndez}},
  \bibinfo{author}{\bibfnamefont{J.}~\bibnamefont{Wintterlin}},
  \bibnamefont{et~al.}, \bibinfo{journal}{Phys. Rev. B}
  \textbf{\bibinfo{volume}{71}}, \bibinfo{pages}{113402}
  (\bibinfo{year}{2005}).

\bibitem[{\citenamefont{Baski and Whitman}(1995)}]{Baski:95}
\bibinfo{author}{\bibfnamefont{A.~A.} \bibnamefont{Baski}} \bibnamefont{and}
  \bibinfo{author}{\bibfnamefont{L.~J.} \bibnamefont{Whitman}},
  \bibinfo{journal}{Phys. Rev. Lett.} \textbf{\bibinfo{volume}{74}},
  \bibinfo{pages}{956} (\bibinfo{year}{1995}).

\bibitem[{\citenamefont{Sze}(1981)}]{Sze:81}
\bibinfo{author}{\bibfnamefont{S.~M.} \bibnamefont{Sze}},
  \emph{\bibinfo{title}{Physics of Semiconductor Devices}}
  (\bibinfo{publisher}{John Wiley and Sons}, \bibinfo{year}{1981}).

\bibitem[{\citenamefont{Snijders et~al.}()\citenamefont{Snijders, Crain,
  McChesney, Gallagher, and Himpsel}}]{unpub:04}
\bibinfo{author}{\bibfnamefont{P.~C.} \bibnamefont{Snijders}},
  \bibinfo{author}{\bibfnamefont{J.~N.} \bibnamefont{Crain}},
  \bibinfo{author}{\bibfnamefont{J.~L.} \bibnamefont{McChesney}},
  \bibinfo{author}{\bibfnamefont{M.~C.} \bibnamefont{Gallagher}},
  \bibnamefont{and} \bibinfo{author}{\bibfnamefont{F.~J.}
  \bibnamefont{Himpsel}}, \bibinfo{note}{unpublished data}.

\bibitem[{\citenamefont{Feenstra et~al.}(1986)\citenamefont{Feenstra, Stroscio,
  and Fein}}]{Feenstra:86}
\bibinfo{author}{\bibfnamefont{R.~M.} \bibnamefont{Feenstra}},
  \bibinfo{author}{\bibfnamefont{J.~A.} \bibnamefont{Stroscio}},
  \bibnamefont{and} \bibinfo{author}{\bibfnamefont{A.~P.} \bibnamefont{Fein}},
  \bibinfo{journal}{Surf. Sci.} \textbf{\bibinfo{volume}{181}},
  \bibinfo{pages}{295} (\bibinfo{year}{1986}).

\bibitem[{\citenamefont{Tersoff and Hamann}(1983)}]{Tersoff:83}
\bibinfo{author}{\bibfnamefont{J.}~\bibnamefont{Tersoff}} \bibnamefont{and}
  \bibinfo{author}{\bibfnamefont{D.~R.} \bibnamefont{Hamann}},
  \bibinfo{journal}{Phys. Rev. Lett.} \textbf{\bibinfo{volume}{50}},
  \bibinfo{pages}{1998} (\bibinfo{year}{1983}).

\bibitem[{\citenamefont{Tersoff and Hamann}(1985)}]{Tersoff:85}
\bibinfo{author}{\bibfnamefont{J.}~\bibnamefont{Tersoff}} \bibnamefont{and}
  \bibinfo{author}{\bibfnamefont{D.~R.} \bibnamefont{Hamann}},
  \bibinfo{journal}{Phys. Rev. B} \textbf{\bibinfo{volume}{31}},
  \bibinfo{pages}{805} (\bibinfo{year}{1985}).

\bibitem[{\citenamefont{Chen et~al.}(1994)\citenamefont{Chen, Wu, Zhang, and
  Lagally}}]{Chen:94}
\bibinfo{author}{\bibfnamefont{X.}~\bibnamefont{Chen}},
  \bibinfo{author}{\bibfnamefont{F.}~\bibnamefont{Wu}},
  \bibinfo{author}{\bibfnamefont{Z.}~\bibnamefont{Zhang}}, \bibnamefont{and}
  \bibinfo{author}{\bibfnamefont{M.~G.} \bibnamefont{Lagally}},
  \bibinfo{journal}{Phys. Rev. Lett.} \textbf{\bibinfo{volume}{73}},
  \bibinfo{pages}{850} (\bibinfo{year}{1994}).

\bibitem[{\citenamefont{Gambardella et~al.}(2002)\citenamefont{Gambardella,
  Dallmeyer, Maiti, Malagoli, Eberhardt, Kern, and Carbone}}]{Gambardella:02}
\bibinfo{author}{\bibfnamefont{P.}~\bibnamefont{Gambardella}},
  \bibinfo{author}{\bibfnamefont{A.}~\bibnamefont{Dallmeyer}},
  \bibinfo{author}{\bibfnamefont{K.}~\bibnamefont{Maiti}},
  \bibinfo{author}{\bibfnamefont{M.~C.} \bibnamefont{Malagoli}},
  \bibinfo{author}{\bibfnamefont{W.}~\bibnamefont{Eberhardt}},
  \bibinfo{author}{\bibfnamefont{K.}~\bibnamefont{Kern}}, \bibnamefont{and}
  \bibinfo{author}{\bibfnamefont{C.}~\bibnamefont{Carbone}},
  \bibinfo{journal}{Nature} \textbf{\bibinfo{volume}{416}},
  \bibinfo{pages}{301} (\bibinfo{year}{2002}).

\end{thebibliography}
\end{document}